\documentclass[acmsmall]{acmart}

\renewcommand\footnotetextcopyrightpermission[1]{}
\newcommand{\revisioncolor}{black}
\usepackage{comment}
\usepackage{tikz}
\usepackage{amsmath}
\usepackage{booktabs}
\usepackage{graphicx}
\usepackage{algorithm}
\usepackage{algpseudocode}
\usepackage{subcaption}
\usepackage{balance}
\usepackage{makecell}
\usepackage{xcolor}
\usepackage{float}  
\usepackage{caption}
\usepackage{wrapfig}
\usepackage{xcolor}
\usepackage{soul}
\usepackage{enumitem}
\usepackage{multirow}
\usepackage{makecell}
\usepackage{graphicx}
\usepackage{subcaption}
\usepackage{makecell}
\usepackage{tabularx}
\usepackage{array}

\newcommand{\code}[1]{\texttt{\small #1}}
\newcommand{\eat}[1]{}

\usepackage{xcolor}
\definecolor{sheshateal}{HTML}{2A9D8F}

\makeatletter
\newcommand{\conextcamreadyfooter}{\footnotesize \@journalNameShort, V\@acmVolume, \@acmNumber, Article \@acmArticle. Publication date: \@acmPubDate.}
\def\@specialsection#1{%
  \let\@vspace\@vspace@orig
  \let\@vspacer\@vspacer@orig
  \par\medskip\small\noindent\textbf{#1:} %
  \let\@vspace\@vspace@acm
  \let\@vspacer\@vspacer@acm
}
\AtBeginDocument{%
  \fancypagestyle{firstpagestyle}{%
    \fancyhf{}%
    \renewcommand{\headrulewidth}{\z@}%
    \renewcommand{\footrulewidth}{\z@}%
  }%
  \fancyhf{}%
  \renewcommand{\headrulewidth}{\z@}%
  \renewcommand{\footrulewidth}{\z@}%
}
\makeatother
\begin{document}
\sloppy

\title[Shesha: Opportunistic In-network Acceleration of Async. DRL]{Shesha: Opportunistic In-network Acceleration of Asynchronous Distributed Reinforcement Learning}


\author{Nehal Baganal Krishna}
\email{nehal.baganal-krishna@ikt.uni-hannover.de}
\affiliation{%
  \institution{Leibniz University Hannover}
  \city{Hannover}
  \country{Germany}
}

\author{Anam Tahir }
\email{anam.tahir@ikt.uni-hannover.de}
\affiliation{%
  \institution{Leibniz University Hannover}
  \city{Hannover}
  \country{Germany}
}

\author{Firas Khamis}
\email{firas.khamis@stud.uni-due.de}
\affiliation{%
  \institution{University of Duisburg-Essen}
  \city{Essen}
  \country{Germany}
}
\author{Mina Tahmasbi Arashloo}
\email{mina.arashloo@uwaterloo.ca}
\affiliation{%
  \institution{University of Waterloo}
  \city{Waterloo}
  \country{Canada}
}

\author{Michael Zink}
\email{zink@ecs.umass.edu}
\affiliation{%
  \institution{University of Massachusetts Amherst}
  \city{Amherst}
  \country{USA}
}

\author{Amr Rizk}
\email{amr.rizk@ikt.uni-hannover.de}
\affiliation{%
  \institution{Leibniz University Hannover}
  \city{Hannover}
  \country{Germany}
}
\renewcommand{\shortauthors}{Baganal Krishna et al.}


\begin{abstract}
Large-scale training for distributed Machine Learning can cause congestion at bottleneck switch ports, leading to model staleness through update losses.
This is particularly detrimental for asynchronous Distributed Reinforcement Learning (DRL) training, as stale updates are known to degrade convergence performance in asynchronous settings.
This paper presents \textit{Shesha}, an in-network DRL accelerator engine, which \textit{opportunistically aggregates} asynchronously generated model updates \textit{on the fly} while they traverse the data plane queue.
This aggregation operation motivates an alternative queue design, which we prototype and envision for future Top-of-Rack switches.
We further present corresponding host-side transmission control in the face of possible congestion, taking advantage of in-network accelerator feedback.
A quantification of model staleness, denoted Age-of-Model (AoM), together with a formal verifier allows us to reason on system-wide AoM objectives in multi DRL-cluster scenarios.
Shesha shows significant reductions in model staleness and queue congestion, improving overall convergence behavior for asynchronous DRL workloads.
\end{abstract}

\maketitle
\makeatletter
\fancyhf{}
\renewcommand{\headrulewidth}{\z@}
\renewcommand{\footrulewidth}{\z@}
\makeatother
\vspace{-5pt}
\section{Introduction}
Reinforcement Learning (RL) is widely used to solve sequential decision problems by learning policies that map observations to actions~\cite{kaelbling1996reinforcement, li2017deep, kober2013reinforcement, xiang2024reinforcement}. Modern RL systems train these policies using repeated cycles of experience collection (through simulations) and gradient-based optimization. As datasets and model architectures grow, RL training has become increasingly distributed: clusters of workers compute local policy updates and periodically transmit them to a central parameter server (PS) or to peer workers for aggregation~\cite{li2014communication, chen2024rina, wan2020rat}.

\begin{figure}[t]
\setlength{\abovecaptionskip}{2pt}
\setlength{\belowcaptionskip}{-10pt}
    \centering
\includegraphics[page=4, width=\linewidth, trim= 0 4cm 0 4cm, clip]{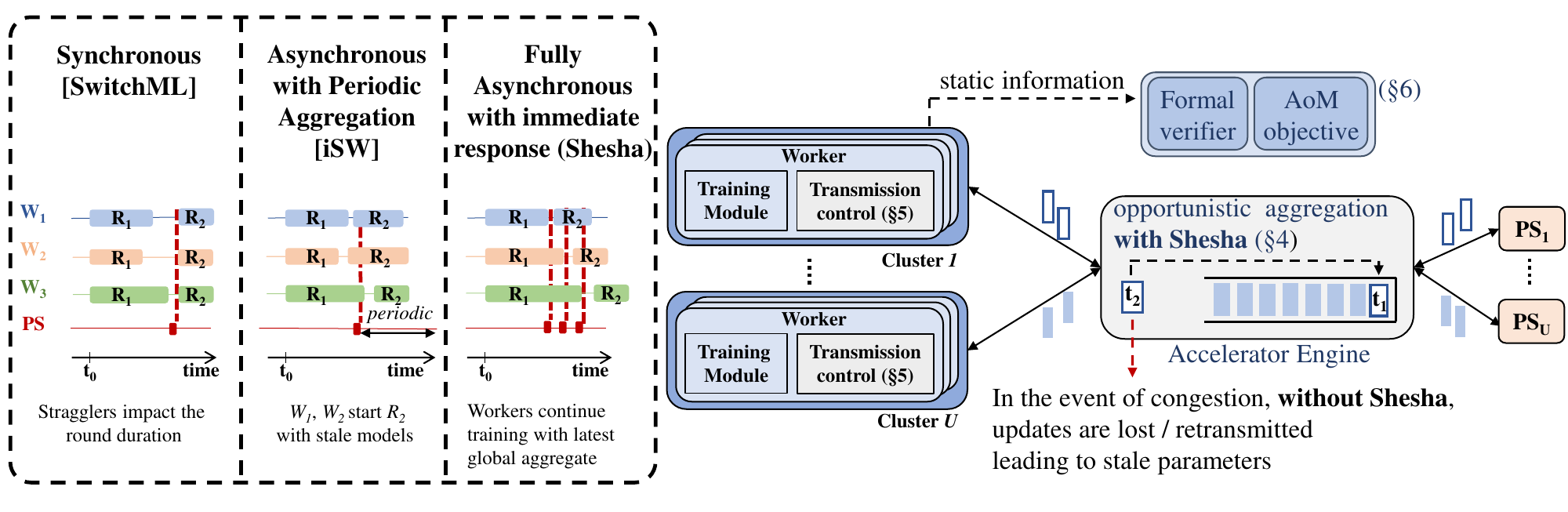}
    \caption{\textbf{(a)} Box on the left: Comparison of update protocols. Horizontal bars $R_i$ represent consecutive training rounds for each worker ($W_1-W_3$). Upon completing a round, each worker directly sends its local update to the parameter server (PS). Red vertical lines indicate the times at which workers receive a global parameter update (red box) from the PS. \textbf{Left:} Synchronous protocol, the PS aggregates only after all workers finish a round, e.g. \cite{sapio2021switchML,lao2021atp}. \textbf{Middle:} Asynchronous protocol with periodic aggregation, the PS updates at fixed intervals using the available updates that arrived, e.g. \cite{iSW}. \textbf{Right:} Immediate-response asynchronous protocol, the PS updates its parameters upon receiving each individual worker update \textit{(Shesha)}. \\\textbf{(b)} Asynchronous DRL training with workers sending updates at different rates. Network congestion events lead to (i) model staleness due to packet loss, (ii) the PS aggregating old model updates (timestamp $t_1$) while other subsuming newer updates ($t_2$) exist. Shesha provides (i) in-network opportunistic aggregation/replacement of DRL model updates, (ii) worker-side transmission control guided by in-network feedback,  and (iii) a formally offline verified model of the update utility expressed through the Age-of-Model metric.}
    \label{fig:side-by-side}
    \vspace{-2mm}
\end{figure}

Large-scale Distributed RL (DRL) predominantly follows either synchronous or asynchronous training. Synchronous methods rely on periodic global barriers to align model updates~\cite{liu2020high,das2016distributed, verbraeken2020survey,li2013parameter, li2024ps}, but suffer from well-known limitations: (i) heterogeneous worker speeds require carefully tuned cutoff timers to close the computation round, (ii) simultaneous update transmissions create network incast, and (iii) slow workers (stragglers) delay global progress.
Fully asynchronous training offers an attractive alternative -- by allowing workers to continuously send updates \emph{without coordination}, it can mitigate straggler bottlenecks, reduce idle time, improve hardware utilization, and naturally accommodate heterogeneity \cite{dean2012large, liu2024asynchronous}. Fig.~\ref{fig:side-by-side}a provides an overview  sketch of these methods.

However, removing synchronization introduces a different set of challenges. In large-scale training tasks, frequent model updates from many workers can congest the network and fill switch queues, leading to update loss and—more subtly—causing workers to train on stale models.
This effect is intensified in multi-tenant environments where multiple DRL training tasks share the same network fabric.
Staleness is particularly harmful in DRL because workers generate updates by evaluating the outcomes of simulations using the \emph{current} model; when this model is outdated, both the collected experience and the resulting gradients reflect obsolete behavior.
Thus, despite its appeal, fully asynchronous DRL remains difficult to scale without addressing these network-induced sources of staleness and loss~\cite{schulman2017proximal, rafailov2023direct, doshi2021distributedPPO,wijmans2019dd}.

%

In this work, we explore how to unleash the benefits of fully asynchronous DRL training using in-network support techniques. 
Specifically, we propose an in-network DRL accelerator engine called \textit{Shesha}, which \textit{opportunistically aggregates} asynchronously generated model updates \textit{on the fly} while they traverse the queue.
When compatible updates -- i.e., updates targeting the same portion of the model or sharing additive gradient semantics -- are present in the accelerator’s queue, Shesha combines them into a single aggregated update.
%
This effectively reduces queue occupancy, significantly reduces packet drops, and maintains the most recent updates in the queue, thereby mitigating staleness and improving end-to-end training efficiency.

As congestion may still arise in the Shesha accelerator engine when many DRL jobs run concurrently, we introduce a worker-side transmission control algorithm that leverages feedback signals provided by the accelerator on the reverse path to regulate the rate at which model updates enter the network.
Workers adjust their transmission rate based on this feedback, preventing the accelerator’s queues from becoming overloaded.
Unlike general-purpose congestion control, this mechanism is tailored specifically to DRL update flows.
In fact, the combination of this transmission control algorithm and asynchronous training is particularly efficient, as the workers can continue simulating and improving their local models while waiting for the next transmission opportunity.

Finally, to quantify model staleness, we introduce \textit{Age-of-Model} (AoM), a metric that captures the utility of a model update, specifically its freshness.
The average AoM of a training task can be related to the average convergence time; in the case of DRL, the time required for the training to achieve a certain reward.
That is, lower AoM corresponds to workers training on more recent models and, consequently, faster convergence.
We provide a formal model of AoM under Shesha’s aggregation and use Satisfiability-Module-Theory- (SMT-) based analysis (e.g., with Z3~\cite{z3}) to statically reason about AoM of \emph{different} concurrent training tasks.
This framework allows us to relate properties such as the maximum update transmission rate to a \textit{system-wide AoM objective} such as verifying that sharing the network does not massively degrade the current workload AoM.


In-network aggregation has been explored in prior systems that accelerate ML and RL training~\cite{yang2022trio, iSW,sapio2021switchML}, but these designs fundamentally rely on synchronous aggregation and assume updates arrive in lockstep rounds, with aggregation scheduled per iteration (e.g., SwitchML~\cite{sapio2021switchML}) or in periodic, synchronized global boundaries (e.g., ATP~\cite{lao2021atp}) (cf. Figure~\ref{fig:side-by-side}a).
Even systems that target asynchronous DRL training still employ periodic synchronization to structure aggregation~\cite{xiao2022asynchronous}.
These approaches are not designed to handle the fine-grained, uncoordinated update streams produced by \textit{fully asynchronous DRL}, where updates arrive continuously and without iteration boundaries. 
For effective aggregation in this setting, the system would need to do so opportunistically, coupled with effective worker-side rate control. Supporting these asynchronous semantics would require fundamental changes to the architecture of existing systems, especially the queues and control protocol.
In contrast, Shesha is designed specifically for such fully asynchronous DRL jobs and, therefore, can perform effective aggregation without imposing synchronization barriers.

\noindent\textbf{Summary of contributions.} We summarize the contributions of this paper as follows: 
\begin{itemize}[leftmargin=20pt]
\item We introduce \textit{on the fly opportunistic aggregation} as an in-network acceleration technique for asynchronous DRL. 
Realizing this technique requires rethinking how packets are queued and forwarded, as well as worker-side transmission control.
As such, we provide an \textit{alternative queue design} with aggregation and replacement primitives for fully asynchronous DRL model updates, along with a worker-side transmission control algorithm guided by the in-network feedback from the accelerator. 
\item We develop this programmable in-network accelerator, denoted  \textit{Shesha}, on a hybrid FPGA-P$4$ pipeline to perform real-time, on the fly opportunistic aggregation \textit{at $100$Gbps line rate}. 
\item We propose the \textit{Age-of-Model} (AoM) as a utility metric to quantify model staleness and provide an offline analysis tool to formally verify system-wide AoM objectives.
\item {\color{\revisioncolor} We evaluate Shesha on a hardware testbed and using ns-3 simulations, showing significant speedup in training time and model update delivery compared to asynchronous with periodic aggregation baselines. }
\end{itemize}

\noindent\textbf{Evaluation Highlights.} {\color{\revisioncolor} We demonstrate, on a live-training hardware testbed with $81$ workers, that Shesha accelerates training, measured through Time-to-Target Reward (TTR\footnote{Similar to Time-to-Accuracy (TTA) in supervised learning, TTR is the time to achieve a target reward for a given workload.}),  compared to iSW-design baselines~\cite{iSW} and a congestion-aware variant iSW-CA, and FIFO across three diverse workloads (\texttt{LunarLander}~\cite{brockman2016openai, towers2024gymnasium}, \texttt{MujoCo HalfCheetah}~\cite{mujoco}, and \texttt{Atari Pong}~\cite{bellemare2013arcade}). 
Shesha is up to $1.5\times$ faster than iSW-CA, $4.7\times$ faster than best iSW configuration, and $8.54\times$ faster than FIFO. Large scale emulations underline this performance showing that Shesha achieves a much lower queueing delay than iSW-design baselines and FIFO under similar aggregation rates.} In a multi-tier topology with heterogeneous, asymmetric capacities, we further observe that Shesha's on-the-fly aggregation cancels out the staleness impairment for the workers affected by local capacity constraints.

\noindent\textit{This work does not involve any data or procedures that present ethical concerns. The paper was entirely produced by the authors without the use of GenAI tools.}

\section{Background} 
\label{sec:background}
\subsection{The Case for Asynchronous Training}
\label{subsec:DRL}
Current Reinforcement Learning (RL) methods to compute optimal policies alternate between simulations to obtain samples, also denoted collected experience, of the currently given policy and gradient-based optimization using objective functions to improve that policy.
%
Updates of the local policy, that we denote \textit{model updates}, are collected, merged, and distributed either by a central entity (parameter server~\cite{li2014communication}) or using an overlay topology (e.g., a ring~\cite{chen2024rina, wan2020rat}) over the network.

Specifically, given $N$ workers, a worker $k$ at step $n$ uses its current local policy $\pi_{x_n^k}$ with parameters $x_n^k$ to collect experience and to calculate parameter gradients $g_k^x$ for the local model parameters $x_n^k$. 
For brevity, we will drop the superscript $(\cdot)^x$ in the sequel.
An objective function $L(x)$ is used to optimize the policy.
Decentralized Distributed Proximal Policy Optimization (DD-PPO)~\cite{wijmans2019dd} builds on the prevalent policy gradient method (PPO). It computes a global model update as
$x_{n+1}^k = \text{GD}\left(x_{n}^k,\text{AllReduce}\left(g_1 L(x_n^1),\dots,g_N L(x_n^N)\right)\right)$, where $\text{GD}$ denotes some gradient descent method~\cite{amari1993backpropagation} and $\text{AllReduce}\left(\cdot,\dots,\cdot\right)$ denotes an \textit{aggregation and broadcast} operation over the input variables. 

%
We distinguish between \textit{\textbf{synchronous training}}~\cite{wijmans2019dd,sapio2021switchML} and \textbf{\textit{asynchronous training with periodic aggregation}} \cite{xiao2022asynchronous,iSW} in DRL. In synchronous training, workers operate in rounds where each worker sends its update to the PS, which waits for all updates, aggregates them, and broadcasts the global model before the next round can begin. 
In \textit{asynchronous training with periodic aggregation}, workers continue training without waiting for the global model. The PS performs aggregation in fixed intervals or after collecting a batch of updates.
Simulation and optimization can run individually on different workers or together on the same worker.

Synchronous training suffers from fundamental drawbacks: (i) coordinating heterogeneous workers is challenging as it involves tuning cutoff timers to close the computation round, (ii) network incast occurs as many workers transmit updates simultaneously, and (iii) stragglers delay global convergence. Asynchronous training with periodic aggregation mitigates straggler bottlenecks, but it introduces staleness, as workers often continue training on outdated models.
To address both the barriers of synchronous training and the staleness of periodic asynchronous schemes, we adopt an \textbf{immediate-response asynchronous} protocol. In this protocol, workers independently transmit updates together with their reward, i.e., utility, and the PS immediately responds with an updated global model whenever the received update improves the global reward. See Fig.~\ref{fig:side-by-side} for examples of the three schemes and their impact on training. A baseline comparison with related work is given in \S\ref{sec:live_training}. For consistency, we refer in this paper to this immediate-response protocol as \textbf{asynchronous training}.

\subsection{Quantifying Staleness by Age-of-Model}
\label{subsec:AoM}

Assuming a large-scale fully asynchronous DRL training mode, model staleness at the aggregation function at the PS describes the passage of time since the latest update from any worker in the cluster. {A \textit{cluster} refers to a group of worker nodes, each independently training a local model and synchronizing through the PS or an in-network mechanism.}
As this global model at the PS is distributed to the workers upon every update aggregation, the staleness of this model is also the staleness of the model used for asynchronous training at the workers. Network congestion leading to update drops and retransmissions, thus lowering the update rate at the PS and to workers asynchronously training with old models. 
The underpinnings of this concept stem from the notion of Age and Value of Information (AoI/VoI), which describe the staleness of information or the timeliness value of updates at a receiver, respectively~\cite{yates2021age}.


\begin{figure}[t]  
\setlength{\abovecaptionskip}{3pt}
\includegraphics[page=12, width=0.6\linewidth,trim=2cm 5cm 0cm 3cm, clip]{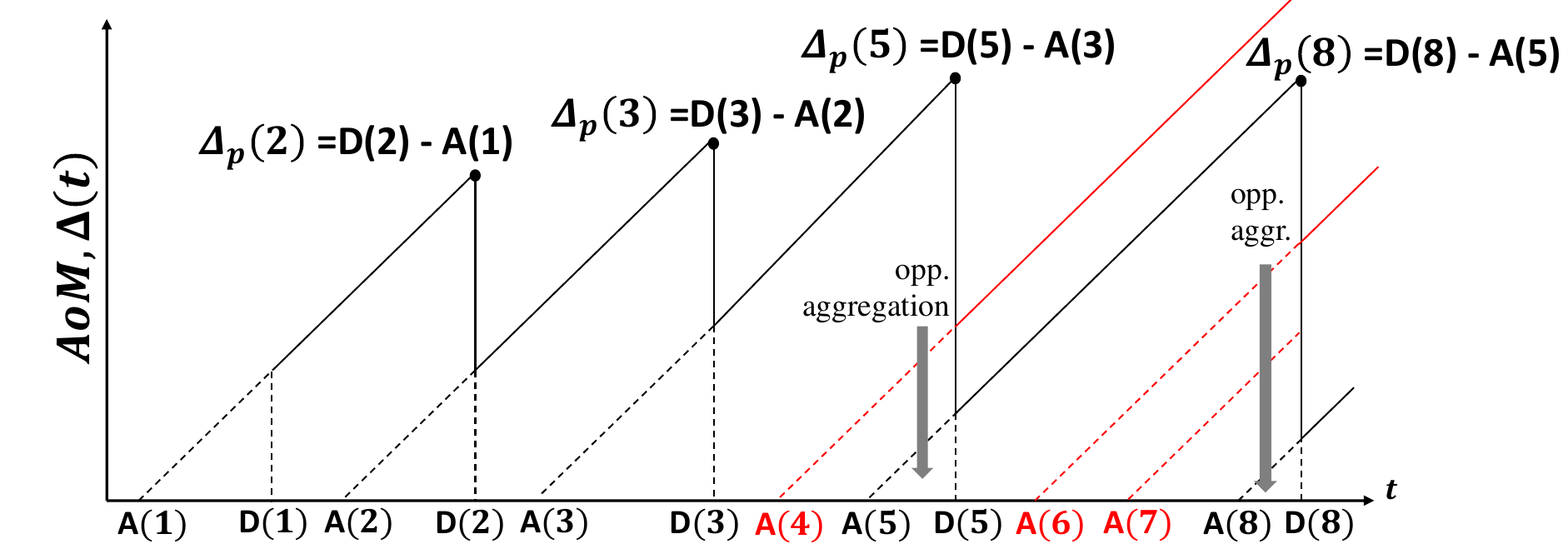}
    \caption{Graphical example of Age-of-Model: The AoM captures the staleness of (global) model at the PS as a function of the \textit{delay and the update transmission times}. $A(n)$ is the update send time, and $D(n)$ is the time at which it (original, aggregated, or replaced) leaves the accelerator engine (see \S\ref{sec:formal_verification}). After aggregation or replacement, models are less stale (less old) and, hence, minimize AoM. Replaced updates are shown in red.}
    \label{fig:aoi_wave}
    \vspace{-5mm}
\end{figure}

AoM defines the staleness (or interchangeably freshness) of a model update as the time that has passed at the PS since the reception of the latest model update.
Note that when a model update from a worker is generated, it obsoletes any previous update from the same worker in the sense that it subsumes the experience collected by the previous model update.
Fig.~\ref{fig:aoi_wave} shows that the AoM is a semi-continuous "sawtooth" function growing linearly with time (i.e. \textit{aging}) and jumping on the event of receiving a model update to the age of that update, i.e., \textit{how much time has passed since this update was created}.
The AoM not only depends on the delay between the worker and the PS but also on the generation and transmission times at the worker. 

\noindent\textbf{Why AoM?} As the AoM provides the age of the updates, we can use it as a proxy metric, first, for the average time required to achieve a specific reward, i.e., given a fixed number of required model updates to achieve a certain reward. A lower average age implies that these updates were received in an overall shorter time span. 
Second, beyond its average, the AoM function reveals through its peak values the times at which the large-scale training application is not performing well, either due to network congestion or slow worker-side update generation, or a combination of both. 
Finally, Shesha uses the AoM to reason about multi-tenant settings, where a formal verification module (cf. Fig.~\ref{fig:side-by-side}b) checks whether the worker-side update transmission parameters lead to large AoM disparities between clusters when they share the network.

\section{Shesha}
\label{Shesha_overview}
Fig.~\ref{fig:side-by-side}b shows an overview of the Shesha system: an in-network acceleration engine, a worker-side transmission control function, and a formal model for statically verifying AoM objectives.

\begin{figure}[t]
\setlength{\abovecaptionskip}{4pt}
\setlength{\belowcaptionskip}{-15pt}
    \centering
    \begin{subfigure}[b]
    {0.48\linewidth}
    \setlength{\belowcaptionskip}{-6pt}
        \centering
        \includegraphics[page=11,width=\linewidth,trim=7cm 7cm 7cm 5cm, clip]{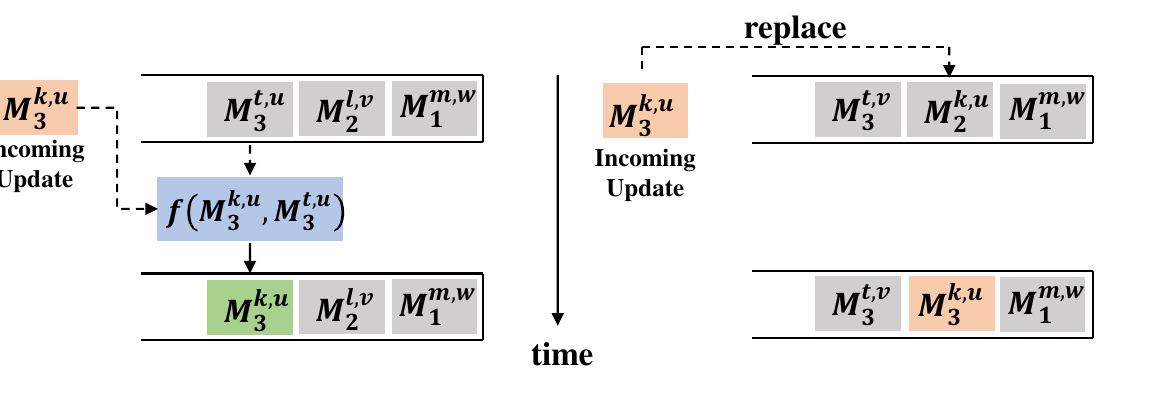}
        \caption{}
        \label{fig:Shesha-primitives}
    \end{subfigure}
    \begin{subfigure}[b]{0.48\linewidth}
    \setlength{\belowcaptionskip}{-6pt}
        \centering
        \includegraphics[page=10,width=\linewidth,trim=5cm 8cm 10cm 2cm, clip]{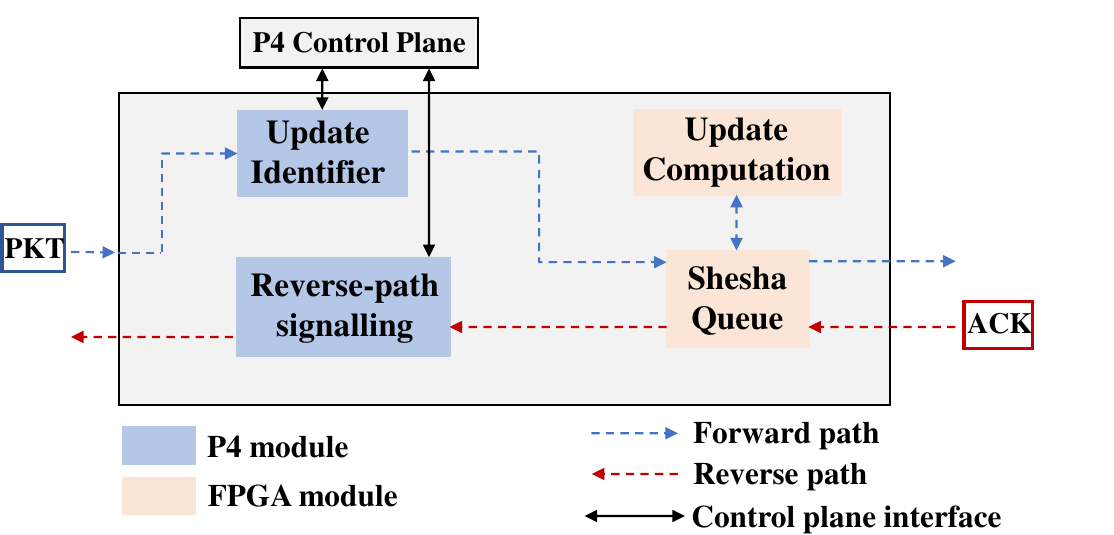}
        \caption{}
        \label{fig:accelerator_engine}
    \end{subfigure}
    \caption{\textbf{(a)} Shesha Queuing Operations: (Left) By default, updates from the same cluster are aggregated. (Right) As a special case, newer updates replace older, unaggregated ones only if they are \textit{from the same worker} to ensure freshness. $M_n^{k,u}$ is the update with timestamp $n$ from worker $k$ that belongs to cluster $u$.\\ \textbf{(b)} Shesha's Accelerator Engine, showing the flow of updates and ACKs through P$4$ and FPGA modules. }
    \label{fig:Shesha}
\end{figure}

\noindent\textbf{Opportunistic Update Aggregation on the Accelerator.} 
Shesha's in-network acceleration engine provides an alternative queue design for on the fly DRL update aggregation. It aggregates model updates in the queue based on the current queue content. 
That is, when the update $M_n^{k,u}$ with timestamp $n$ from worker $k$ in cluster $u$ (representing either the entire model gradient values $g_k$ from §~\ref{sec:background} or a segment of that, see below) arrives at the Shesha queue, the acceleration engine checks if there are other updates from the same cluster in the queue. If yes, $M_n^{k,u}$ is merged with the waiting update $M_{m}^{l,u}$, and the result is written back to the queue at the position where the waiting update resides. If there is no other update from the cluster waiting,  $M_n^{k,u}$ is placed at the end of the queue.
%
This opportunistic aggregation, illustrated in Fig. \ref{fig:Shesha-primitives}, ensures that \emph{at most one update per cluster} is in the queue at any point in time.

\noindent\textbf{Opportunistic Update Replacement.} 
A special case is updates from a worker who only finds an older update from the same worker waiting. 
In some asynchronous modes, such as Multi-agent distributed RL in~\cite{iSW}, workers can inject multiple updates before receiving a new aggregate model/policy.
A key observation is that the more recent update \textit{from the same worker}, with a smaller AoM, subsumes the older update as it is computed using more recent experience.
Hence, Shesha's acceleration engine replaces the older update in the queue with the new one. 
%
Note that model aggregation resets the subsuming property of subsequent updates from the same worker, i.e., \textit{replacement occurs if and only if two unaggregated models of the same worker meet in the queue}.

In aggregating updates, Shesha is careful to do so in a convergence-preserving manner by filtering low-reward updates. That is, when a new update arrives, and another update from the same cluster is in the queue, Shesha compares the reward value in the incoming update with the reward in the queued one. If the rewards are comparable, i.e., the difference is within a given threshold, the gradients are aggregated. If the incoming update has a higher reward surpassing the threshold, it replaces the existing one. Conversely, if the reward is significantly lower, Shesha drops the update. This mechanism ensures the queue retains convergence-preserving updates, which we empirically observed to lead to stable convergence in \S~\ref{sec:live_training}.

Note that because Shesha targets fully asynchronous DRL, the accelerator does not wait to collect a fixed number of updates nor enforce any form of round synchronization.
That is, it \emph{does not} hold an update longer than it would have waited in a normal queue.
Instead, it aggregates updates opportunistically  whenever multiple updates of the same cluster are arriving/waiting in the queue. 
In case many workers simultaneously send their updates, Shesha naturally \textit{mitigates incast by compressing all updates of the same  cluster in the queue into a single model update}, reducing congestion and load without sacrificing update quality.  

\noindent\textbf{Model Updates spanning Multiple Packets.} A model update $M_n^{k,u}$ from worker $k$ can represent either the full gradient values $g_k$ or a segment of that. 
%
When a model update does not fit into one packet, Shesha, similar to prior in-network aggregation systems~\cite{sapio2021switchML}, has workers divide gradients into multiple, consistently indexed update segments (i.e., MTUs) along fixed boundaries.
Each packet then carries a distinct, deterministically numbered portion of the gradient.
Aggregation is then performed only between updates from the same cluster that correspond to the same segment. Unless stated otherwise, we assume for simplicity that each update represents the entire model gradient. The implications and handling of model segmentation are discussed where relevant.  

\noindent\textbf{Transmission Control Function.} In fully asynchronous DRL, workers may transmit updates to the PS immediately after generating a fresh update. 
Without regulation, simultaneous transmissions from many workers across a large number of different clusters can overwhelm shared links and cause congestion~\cite{wang2020domain,chen2023boosting}.
To prevent this, Shesha workers adapt their transmission rate based on the network conditions. ACKs returning from the PS traverse the accelerator engine and are annotated with the current occupancy of the bottleneck queue (Shesha assumes symmetric forward and reverse paths, as is typical when deployed in a ToR switch).
Hence, the worker-side transmission control accurately infers congestion at the accelerator and makes informed decisions on when to compute and send an update. When the queue is busy, workers may continue collecting experience rather than injecting traffic; when the queue is light, they can transmit promptly. This closed-loop mechanism helps avoid congestion, reduces packet loss, and minimizes the average AoM.

\noindent\textbf{AoM Formal Model and Objective Verification.} 
The rate at which each cluster worker transmits updates influences the cluster's share of the bottleneck link.
If worker transmission-control parameters are not set carefully, it may lead to disparities between clusters in update delivery rates at the PS and hence disparities of cluster AoMs.
%
Shesha provides a formal model of the AoM that takes the parameters of the transmission control functions to verify system-wide AoM objectives over clusters. 
{\color{\revisioncolor}In \S\ref{sec:formal_verification}, we analyze the worker-side transmission control with the aim of bounding the AoM disparity caused by sharing the network.}

\section{In-Network Accelerator Hardware Architecture}
\label{sec:in_network_engine}

This section describes how Shesha's in-network accelerator (Fig.~\ref{fig:accelerator_engine}) performs opportunistic aggregation of DRL updates from asynchronous workers in the data plane.
Additional details about the hardware implementation are given in \S\ref{subsec:hardware_extra}. 

To decide if an arriving update should be aggregated, replace an older one, or simply enqueued, the accelerator engine must distinguish updates by their origin.
Shesha workers tag each update with a \code{Worker\_ID} and a \code{Cluster\_ID}, whose combination uniquely identifies each worker's updates.
For segmented models, a \code{Segment\_ID} is added to uniquely identify updates corresponding to different gradient segments.
The accelerator's \code{Update Identifier} (Fig.~\ref{fig:accelerator_engine}) extracts these identifiers from incoming updates, and forwards them, along with the update itself, to \code{SheshaQueue} for processing. 
%


\noindent\textbf{SheshaQueue.} The Shesha queue holds the model updates waiting to be transmitted on the bottleneck link. 
The enqueue logic depends on the current queue content. When enqueuing a new update, the accelerator engine must search for and locate any existing update in the queue that matches the combinations of the \code{Cluster\_ID}, \code{Worker\_ID}, and \code{Segment\_ID} of the incoming update.
The result of the search
determines if the incoming update is aggregated, replaces an older one, or is appended to the queue tail.
To locate relevant queued updates efficiently, we design the \texttt{SheshaQueue} to track and modify updates as they traverse through it, as we discuss below.
Dequeuing occurs sequentially from the queue head.

\begin{figure}[t]    
\setlength{\abovecaptionskip}{1pt}
\setlength{\belowcaptionskip}{-10pt}
    \includegraphics[page=16, width=\linewidth, trim=0cm 11cm 0cm 1cm, clip]{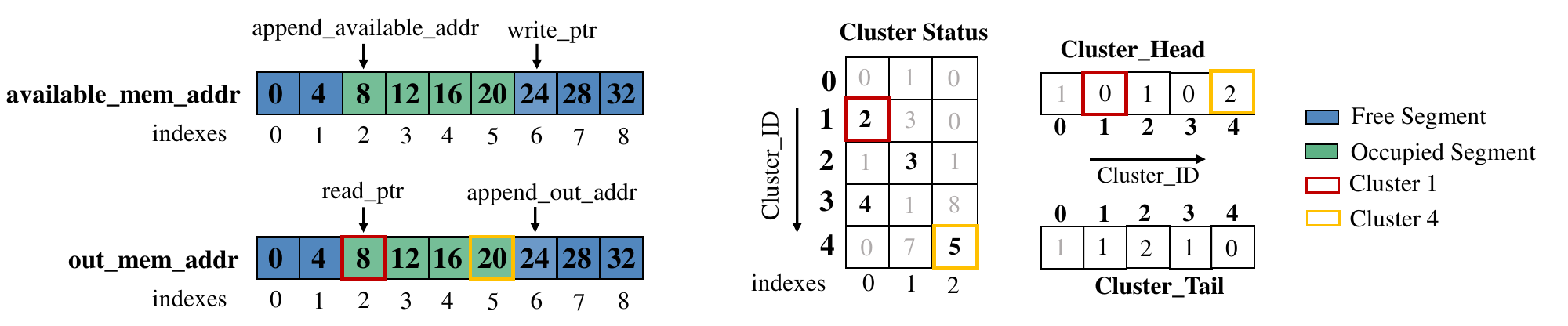}
    \caption{ Per-cluster update tracking: Each segment is made up of $4$ memory blocks; therefore, the segment addresses in the lists on the left are multiples of $4$. Shesha uses \texttt{cluster\_status}, \texttt{cluster\_head}, and \texttt{cluster\_tail} to track update addresses per cluster; if head and tail are equal, as in Cluster $0$, no update is queued. \texttt{cluster\_head} points to the index in the cluster's ``row'' in \texttt{cluster\_status}, which holds the queue index of that cluster's first update. This index is then used to fetch the corresponding segment address from \texttt{out\_mem\_addr} for transmission (Clusters $1$ and $4$). }
    \vspace{-10pt}
    \label{fig:overview}
\end{figure}

\noindent\textbf{Memory Organization}. 
The queue's memory is divided into fixed-size segments with pre-determined address boundaries, each capable of storing one uniformly sized update, forming the fundamental blocks for handling queued updates.
In the example in Fig.~\ref{fig:overview}, each segment is made up of four consecutive memory blocks, and segment addresses are multiples of 4. 
%

\noindent\textbf{Data Structures.} Four data structures help track and process the queued updates. \code{available\_mem\_addrs}tracks the empty segment addresses where an incoming update can be stored, while \code{out\_mem\_addrs} tracks segment addresses of queued updates in sequential order. 
\
We use cyclic pointers over these lists during enqueue and dequeue. 
\code{write\_ptr} and \code{append\_out\_addr} are updated during enqueue to ``move'' the next available segment from \code{available\_mem\_addrs} to \code{out\_mem\_addrs}.
\texttt{read\_ptr} and \texttt{append\_available\_addr} are updated similarly during dequeue (Fig.~\ref{fig:overview} with further details in \S\ref{subsec:hardware_extra}).
For each \code{Cluster\_ID}, \code{cluster\_status} stores the pointer to that cluster's update in the queue -- for segmented models, the data structure stores this per cluster and segment pair. Finally, \code{replace\_status} tracks if a cluster holds a non-aggregated replaceable update using one-bit \code{replace\_flag} and a corresponding \code{Worker\_ID} to allow update replacement from the same worker.



\noindent\textbf{Enqueue Process.} During the enqueue process for an update with identifiers \code{Cluster\_ID} and \code{Worker\_ID}, the accelerator checks \code{cluster\_status} and \code{replace\_status} to see if there is an existing update for \code{Cluster\_ID} and if it is replaceable (i.e., not aggregated and from the same worker).
If so, the accelerator uses the pointer \code{cluster\_status} along with \code{out\_mem\_addrs} to accurately locate the existing update and replace it by overwriting the queue memory. 
If not replaceable, it aggregates and updates \code{cluster\_status} to disable further replacement.  

If there are no existing updates, the accelerator adds the pointer pointing to the segment where the update is being stored in \code{cluster\_status}, and sets \code{replace\_status} to indicate that the cluster contains a replaceable update from a worker, along with the associated Worker\_ID. 
The incoming update is only dropped if the queue is full \emph{and} no update for the same cluster is present.
Further modifications for segmented model updates are illustrated in \S~\ref{sec:imp}.

\noindent\textbf{Dequeue Process.} Departures are strictly sequential. Without aggregation or replacement, the dequeuing process is FIFO. If an update is aggregated or replaced, the resulting update retains the original update’s position in the departure order.
\section{Worker Transmission Control}
\label{sec:worker_algo}
The data plane acceleration engine described in §\ref{sec:in_network_engine} is independent of the worker-side update transmission control algorithm. 
In this section, we describe the rationale behind the worker-side update transmission rate control that arises through the alternative queue design in the Shesha queue. We then describe a feedback mechanism from the accelerator engine to the workers and finally worker-side transmission control to handle congestion in the queue.

\noindent\textbf{How should a Worker pick a Transmission Rate?} 
Given that the goal of the DRL training application is to minimize staleness, i.e., AoM, the workers need to pick a transmission rate that keeps congestion low. 
The alternative queue design of Shesha, specifically, the opportunistic aggregation per cluster, implies that congestion only arises with Shesha when the number of active clusters $U$ exceeds the number of queue memory slots (queue size) $Q_{\max}$. 
This is due to the queue keeping at most one update at a time from every cluster, and each update exactly requires one memory segment. 
Respectively, this arises under model segmentation when the number of clusters $U$ times the number of segments $\Tilde{S}$ exceeds $Q_{\max}$. 
Hence, as long as $U\cdot \Tilde{S} \leq Q_{\max}$, the workers can transmit updates at the highest possible rate, which also minimizes the AoM.
Otherwise, the workers have to pick a transmission rate that collectively reduces congestion, hence reducing the update losses affecting the AoM. 
In the following, we assume $\Tilde{S}=1$ for notation clarity and propose a feedback-based transmission control algorithm that acts as an arbitration scheme.

\noindent\textbf{Reverse-path Signaling.} Workers can make informed decisions about whether to transmit model updates, if they have information about the recent status of the queue. 
Specifically, knowing if there is congestion or not (the number of active clusters being higher or less than the queue size) can guide the workers to control their transmission rate. 
Inspired by Reverse-path Congestion Marking
~\cite{feldmann2019p4,baganal2023rpm}, 
we leverage ACKs on their way back to workers to additionally carry queue status updates (see Fig.~\ref{fig:accelerator_engine}). {\color{\revisioncolor}Shesha uses this signaling to drive AoM-aware update transmission rate adjustment at the workers.}
This approach provides workers with recent \texttt{SheshaQueue} congestion status while avoiding additional status update traffic in the system. 

\noindent\textbf{Worker-side Protocol.} For now, we consider the topology in Fig.~\ref{fig:side-by-side} with the accelerator residing in the middle between $U$ active clusters and a number of PS. Assume the queue has a memory  of size $Q_{\max}$ and that the current utilization of that memory is $Q_n$ at time $n$. 
The accelerator engine piggybacks a signal on the ACKs delivered to the workers, specifically, the number of currently active clusters sending updates and the queue utilization. 
Hence, the ACK delivers the queue state $\{U, Q_{max}, Q_n\}$ to the workers. 
As $Q_{max}$ is static, it need not be transmitted after the initial time, and to detect a full queue, Shesha can also send a binary variable instead of transmitting $Q_n$. 
To gauge worker-side decisions on the timeliness of the feedback received, each worker $k$ \textit{keeps a timer} $\hat{\Delta}_{k,n}$, which is the time since the last ACK it received. 

Now, we tie the transmission urgency to the ability of the DRL worker-side application to tolerate delayed feedback. 
When $Q_{\max} < U$, i.e. in the congestion regime,  and worker~$k$ has a new model update to send, it decides to transmit the update with probability $P_s = \min\left(\frac{Q_{max}}{U} + f(\hat{\Delta}_{k,n}),1\right)$. 
Here, $f(\hat{\Delta}_{k,n})$ is homogeneous among workers and captures the ability of the DRL worker-side application to tolerate delayed feedback. 
Specifically, we fix a timer threshold $\bar{\Delta}_{T}$ such that if the time since the last received ACK $\hat{\Delta}_{k,n} > \bar{\Delta}_{T}$, the worker considers the latest received queue state obsolete. We set $f(\hat{\Delta}_{k,n}) = v \cdot (\hat{\Delta}_{k,n} - \bar{\Delta}_{T})$ if $\hat{\Delta}_{k,n} - \bar{\Delta}_{T}>0$ and to zero otherwise. 
Hence, all workers with non-obsolete feedback transmit their updates with the fixed probability $P_s = \frac{Q_{max}}{U}$ leading to stability. 
Workers with obsolete feedback perturb this transmission probability with $f(\hat{\Delta}_{k,n})\geq 0$. 
{\color{\revisioncolor}The slope $v>0$ in $f(\hat{\Delta}_{k,n})$, controls how aggressively workers increase their transmit probability once feedback becomes stale. We choose $\bar{\Delta}_T$ and $\beta$ so that sharing the bottleneck does not substantially increase any cluster's AoM or create a large AoM degradation gap.}

\section{Verifying Age-of-Model Objectives}
\label{sec:formal_verification}

Staleness reduction is a major driver behind Shesha. As such, we study staleness across clusters that concurrently share a Shesha queue using Shesha's final component (\S\ref{Shesha_overview}), 
%
a static formal verifier that allows reasoning about staleness, i.e., AoM objectives, across clusters. 
The static verifier takes in (1) a formal model, inspired from \cite{cacc_mina} and represented as a set of constraints encoding how AoM changes over time as updates are sent over the network, (2) parameters of the clusters, and (3) a query on an objective function of the resulting AoM functions of all clusters, 
%
and returns a binary result on whether the parameterized model satisfies the objective. 
We show how this supports static reasoning about AoM {\color{\revisioncolor}degradation} between clusters. We envision this to be used for constraining the cluster parameters to adhere to given AoM objectives.  

\noindent\textbf{A Formal Model of AoM.} 
Fig.~\ref{fig:aoi_wave} shows how AoM evolves in the presence of Shesha's in-network aggregation and replacement.
Consider multiple workers of one cluster sending updates through the accelerator engine. $A(m)$ denotes the time at which an update $m$ from any of the workers arrives at the accelerator, and $D(m)$ the time at which this update leaves the accelerator queue. Corresponding times when any worker has generated the $m$th model update at time $A_1(m)$ and when it was received by PS at time $D_1(m)$ are shifted by a constant transmission delay, which we do not consider here for simplicity.
The  difference $D(m)-A(m)$ is the one-way delay for delivering that update.
%
We assume that the PS transmits the ACK back to the cluster immediately upon receiving the update, and that the ACKs reach the cluster after a constant delay $\delta$.
The AoM $\Delta(t)$ at the PS at time~$t$ determines the age of the last update received at the PS, and hence how old the global model distributed to the asynchronous workers is.
Note that aggregation leads to lower AoM, which means fresher, i.e., less stale, model updates.
We consider the peak Age-of-Model $\Delta_p$, which is the maximal AoM just before a valid model leaves the accelerator engine. 
The peak AoM $\Delta_p(m)$ after update $m$ is received is 
$\Delta_p(m) = (D(m) - A(m^\prime)) \cdot 1_{\left\{D(m) < A(m+1)\right\}} $
where $m^\prime$ is the index of the latest departed update, $m^\prime = \max \{ i < m: D(i) < A(i+1)\}$ with indicator function  $1_{\left\{X\right\}}$.


\noindent\textbf{Analyzing Worker-side Transmission Parameters.} 
We use Satisfiability Modulo Theories (SMT) to reason about queries expressed as a logical first-order formula over the AoM function.
We can formally verify the behavior of the transmission control parameters of the workers by reformulating the AoM model into a series of logical expressions (details in the appendix in §~\ref{subsec:AoM_details}).
Based on these expressions, the verifier can deem parameters of the transmission control per cluster to either fulfill a global logical condition on model staleness, i.e., the AoM objective, or not, and, hence, accept or reject a cluster configuration. 

\noindent\textbf{AoM {\color{\revisioncolor}Degradation} as Objective.} 
{\color{\revisioncolor} We define AoM degradation as the AoM increase caused by sharing the network with other clusters. We apply the formal model to verify whether the selected parameters of the per-cluster transmission control keep this degradation below the bound $\Gamma$ and limit this degradation across clusters. 
The parameters include the threshold $\bar{\Delta}_{T}$ in §\ref{sec:worker_algo}, which decides on the base backoff time for the workers upon congestion.
The corresponding  constraints in the formal model are provided in \S\ref{subsec:AoM_details} and are encoded together withonstraints given in Appendix~\ref{subsec:AoM_details} are encoded together with bounded workload-generation assumptions, standalone AoM baselines, and candidate worker-side transmission-control parameters. 
The solver derives admissible update arrival and departure times that satisfy these constraints and checks whether any bounded trace violates the AoM degradation objective. 
Specifically, we query the SMT solver for the existence of a counterexample trace in which at least one cluster's AoM under sharing the network exceeds its standalone AoM baseline by more than the target degradation bound (cf. \S\ref{sec:multihop_ns3}).  
}

\section{Accelerator Implementation}
\label{sec:imp}

We have developed a prototype Shesha's accelerator engine on an FPGA-based accelerator card~\cite{amdU55C} that can operate at $100$Gbps line rate. We envision that such a prototypical implementation can be integrated into future ToR switches that combine switching fabrics and different hardware acceleration platforms~\cite{cisco8100}. For our evaluation, we describe how we use this prototype with a Tofino switch in our testbed in \S\ref{sec:eval}.

\texttt{SheshaQueue}, which consists of all queuing-related operations, is written in Verilog. The Update Identifier (Fig.~\ref{fig:accelerator_engine}), which extracts the \code{Cluster\_ID} and \code{Worker\_ID} from the incoming updates and embeds the queue status in the ACKs, is implemented in P$4$~\cite{bosshart2014p4}. 
This combination of Verilog and P$4$ 
leverages the strengths of both languages harmoniously. P$4$'s parsing, deparsing, and lookup engines make it a natural choice for identifying updates and embedding queue status in ACKs, as these tasks involve header parsing, using match-action tables to extract IDs, and deparsing the update to embed queue status in ACKs. Verilog allows fine-grained hardware specification of the Shesha queue and its operations.

\noindent\textbf{Packet Format.} 
Packets include $304$ bits of standard connectionless header fields on top of Ethernet, followed by the mean reward of local iteration, the gradients, and {\color{\revisioncolor}the aggregation count, which records how many worker gradients are combined in the packet.}
ACKs on their way back to the workers contain the aggregated gradients. The reverse-path signaling adds the queue status fields, up to $24$-bit queue utilization, and $16$-bit \# of active clusters.

\noindent\textbf{FPGA Implementation Details and Resource Footprint.} 
We prototype Shesha on Alveo U55C accelerator cards~\cite{amdU55C}, which contains the Virtex XCU55 UltraScale+ FPGA. 
Shesha incurs a minimal hardware footprint (details in \S~\ref{subsec:fpga_resources}).
We use the AMD OpenNIC shell~\cite{xilinxOpenNIC}, an FPGA-based NIC framework that provides a high-speed, modular networking platform with well-defined data and control interfaces for seamless integration of custom user logic. It includes core components that implement host and Ethernet interfaces, along with two user logic blocks for integrating Register Transfer Language plugins. 
We place Shesha's Acceleration Engine in the one operating at $250$ MHz.
We convert the P$4$ programs into FPGA-compatible modules using Vitis Networking P$4$ (\texttt{VNP4})~\cite{amdVitisNetP4}.

\noindent\textbf{Packet Processing Pipeline} The Shesha pipeline within the User Logic block of the OpenNIC shell consists of three stages: the first P$4$ block (\texttt{VNP4}$_1$), then the Verilog-based \texttt{SheshaQueue}, implemented using hardware registers 
to efficiently support various packet operations/aggregations, and finally the second P$4$ block (\texttt{VNP4}$_2$). These components communicate using the AXI4-Stream protocol without any bridges in between.
%

\eat{
To convert the P$4$ design into an FPGA-compatible module, we use Vitis Networking P$4$ (\texttt{VNP4})~\cite{amdVitisNetP4}, which enables the integration with other IPs over the AXI4-stream protocol. We design \texttt{SheshaQueue} to use the AXI4-stream protocol~\cite{axi4stream} as an interface to communicate with \texttt{VNP4} modules without any bridge in between.
The Shesha pipeline within the User Logic block of the OpenNIC shell consists of three stages: the first P$4$ block (\texttt{VNP4}$_1$), then the Verilog-based \texttt{SheshaQueue}, and finally the second P$4$ block (\texttt{VNP4}$_2$). These components communicate using the AXI4-Stream protocol.
}
 


\noindent\textbf{Uplink (Workers to PS).} \texttt{VNP4}$_1$ receives updates and first checks whether they originate from known worker nodes by matching the source IP. Packets not matching any known worker or the parameter server are treated as non-DRL traffic and bypass all model-related processing. For valid model updates, \texttt{VNP4}$_1$ extracts \code{the Worker\_ID} (computed as a five-tuple hash of the packet) and assigns a \code{Cluster\_ID} by mapping the worker to a multicast group. The extracted metadata (\code{Cluster\_ID} and \code{Worker\_ID}) is forwarded (as AXI user signal) along with the update to the \texttt{SheshaQueue}, which performs opportunistic aggregation. The outgoing update is forwarded unmodified by \texttt{VNP4}$_2$ to the parameter server.

\noindent\textbf{Downlink (PS to Workers).}
On the downlink, \texttt{VNP4}$_1$ processes ACKs identical to the uplink by extracting metadata (Worker\_ID and \code{Cluster\_ID}) and forwarding them to the \texttt{SheshaQueue}. These packets are identified as non-DML packets, and Shesha forwards them without queue interaction. For Reverse-path signaling, \texttt{VNP4}$_2$ embeds queue status received by \texttt{SheshaQueue} (as AXI user signal) in the ACK and multicasts the ACKs to all workers in the corresponding \code{Cluster\_ID} using preconfigured multicast groups. This ensures that all workers contributing to an aggregated update receive the latest global model and queue status.


\noindent\textbf{P$4$ Control Plane.} 
The P$4$ control plane interfaces with the accelerator engine to dynamically configure the match-action tables of the P$4$ blocks. This allows for flexible adaptation to evolving workloads, such as re-grouping workers into existing clusters or forming new clusters on the fly, thereby supporting multi-tenant distributed training jobs. 

\noindent\textbf{Support for Model Segmentation.}
We extend \texttt{VNP4}$_1$ to extract the \code{Segment\_ID} from updates to identify the specific subset of model gradients in the update. \texttt{VNP4}$_1$ then computes a hash value over the \code{Cluster\_ID} and \code{Segment\_ID} to uniquely identify the segment. This replaces the \code{Cluster\_ID} as the primary key to store and track updates, and for aggregation and replacement logic.

\section{Evaluation}
\label{sec:eval}

\textcolor{\revisioncolor}{
We describe our evaluation testbed setup below. 
Our testbed evaluation includes a baseline comparison with approaches that are asynchronous with periodic aggregation in a large scale DRL emulation, specifically iSW and a congestion-aware version we denote as iSW-CA (\S\ref{sec:networkbenchmarks}). We use three different live DRL training workloads to show how Shesha improves the training speed in comparison to various baselines (\S\ref{sec:live_training}), \textit{all on the hardware testbed}.  
Finally, we show the impact of Shesha's worker-side transmission control on the AoM in a large-scale multi-tier scenario using ns-3 simulation (\S\ref{sec:multihop_ns3}).
}

\noindent\textbf{Evaluation and Testbed Setup.}  
\textcolor{\revisioncolor}{The evaluations in \S\ref{sec:networkbenchmarks} - \S\ref{sec:live_training} are carried out on our hardware testbed consisting of Alveo U55C accelerator cards~\cite{amdU55C} and a $32\times100$ Gbps Intel Tofino switch \cite{intelTofino} in addition to $81$ end-hosts used for live DRL training.
Since it is not possible to modify the bottleneck FIFO queue on the ToR Tofino switch \cite{intelTofino}, we deploy the accelerator engine on a bump-in-the-wire U55C Alveo between the switch and the parameter server. 
For the evaluations in \S\ref{sec:networkbenchmarks}, an FPGA-based traffic generator is connected directly to the Alveo running Shesha. It injects controlled high-rate traffic into the SheshaQueue\footnote{The traffic generator replays and scales DRL worker traces to emulate $2000$ workers at aggregate output rate of $100$~Gbps.}. 
For the live training experiments in \S\ref{sec:live_training}, $81$ physical workers train without traffic emulation. 
Their traffic is merged at the Tofino switch, forwarded through Shesha to the parameter server, and the corresponding ACKs and model responses traverse the reverse path back to the workers.} 
That is, the Tofino is only used for switching and forwarding the merged stream to the Alveo card that implements the SheshaQueue operations at the bottleneck link and operates at $100$Gbps line rate.
Depending on the experiment goal and the combined rate of the hosts, we vary Shesha's outgoing rate to analyze the performance under different congestion settings.
All reported metrics are averages over 30 runs. 


{\color{\revisioncolor}
\noindent\textbf{Evaluation Metrics.} For the large scale emulation on the hardware testbed, we report the packet drop rate, the model segment aggregation rate, the model segment aggregation size, and the packet queueing delay. 
For live training, we report the \textbf{Time-to-Target Reward} (TTR) as well as the final reward achieved in training for standard training horizon. 
For multi-tier evaluations, we report AoM degradation (staleness), i.e., the AoM increase for a growing number of training clusters.
}

{\color{\revisioncolor}
\subsection{Large-scale Emulation on the Hardware-Testbed}
\label{sec:networkbenchmarks}

To illustrate Shesha's advantages, we compare its performance against the baselines that do not do in-network aggregation (FIFO) and that deploy asynchronous with periodic aggregation mechanisms (iSW, iSW-CA). 
We emulate $2000$ workers using FPGA-based replay, where each worker sends $200$ model updates of $2.3$~MB each, segmented into $1540$ MTU-sized packets. 
The combined traffic has an aggregate rate of $100$ Gbps. 
Each packet carries a \code{Segment\_ID} (defined in \S\ref{sec:imp}) such that packets with the same \code{Segment\_ID} represent the same model segment from different workers. 
}


\noindent \textcolor{\revisioncolor}{\textbf{Baselines.}
{iSW}~\cite{iSW} proposes to aggregate model segments in switch registers. In its default behavior, the aggregation continues until all workers have submitted their update, which is prone to worker stalling (cf. Alg. 1 in~\cite{iSW}). 
It allows workers that have submitted their update to asynchronously continue training  on the old model for a number of iterations. 
We base our \textit{asynchronous with periodic aggregation} baseline on {iSW}. \textbf{iSW-max} implements iSW's default behavior described above. 
\textbf{iSW} denotes baselines with iSW-style aggregation with configurable time intervals. That is, When an incoming update arrives it is aggregated into its corresponding model segment. 
The switch broadcasts the aggregate back to the workers (denoted here \textit{flush-ready aggregate}) at the end of a fixed batching window (10$\mu$s - 1ms across different experiments). We also design a \textit{congestion-aware} iSW variant denoted \textbf{iSW-CA}, which uses the batching window, however, if the output buffer is full when the window closes, it keeps aggregating into the flush-ready aggregate until there is space in the output buffer. 
We implement SheshaQueue, FIFO, iSW, iSW-max, and iSW-CA on Alveo U55C.
}

{\color{\revisioncolor}
\noindent\textbf{Shesha achieves fewer drops and lower queueing delay.}
Tab.~\ref{tab:combined_output_rate_comparison} compares Shesha against FIFO, iSW and iSW-CA using four metrics: (1) drop rate, i.e., the fraction of incoming packets dropped, (2) queue delay, i.e., the time a packet waits in the system for aggregation and transmission, (3) aggregation rate (\texttt{agg\_rate}), i.e., the fraction of incoming packets merged with another packet, and (4) aggregation size, which measures how many incoming packets are represented on average by a single outgoing packet (roughly $\texttt{agg\_size} \approx 1 + \omega \cdot \texttt{agg\_rate} $). 
Note that iSW-based designs have a fixed aggregation window, and their performance, as we show later in the section (Tab.~\ref{tab:network_isw_isw-ca_out_varied_compact}), depends on the chosen window size.  For each congestion condition in Tab.~\ref{tab:combined_output_rate_comparison}, we picked the window size that results in similar aggregation rate as Shesha. 
Also note that aggregate rate and size are not applicable to FIFO with no in-network aggregation.
Recall that the rate of incoming packets, 
$R_{\mathrm{in}}$, is $100$~Gbps. We sweep different congestion conditions, defined as load factor $\omega := R_{\mathrm{in}}/R_{\mathrm{out}}$  by varying the output rate $R_{\mathrm{out}}$.

Shesha outperforms iSW by avoiding fixed-window batching. iSW is sensitive to tuning, i.e., the batching window may need to be optimized for the specific workload and load factor combination. Small batching windows provide limited aggregation and cause drops under congestion, while large batching windows reduce drops but deliver delayed updates later (cf. Tab.~\ref{tab:network_isw_isw-ca_out_varied_compact}). 
In iSW, the window is fixed, so as network conditions vary, aggregates may leave from the switch aggregators to the output queue too early and get dropped, or may stay too long and are unnecessarily delayed. 
Our extension iSW-CA outperforms iSW and comes closest to Shesha as it extends the aggregation window under congestion and avoids dropping flush-ready aggregates. However, it still uses an aggregation window, and it holds the aggregates until output buffer space is available. As such, Shesha shows lower queuing delay than iSW-CA. FIFO performs the worst as it does no aggregation and loses packets under congestion. As iSW-max waits for all worker updates to aggregate, it achieves, across all $\omega$, an aggregation size of 2000, an aggregation rate of nearly $100$\%, and no drops, but at the cost of high queueing delay that is three orders of magnitude higher ($249$ms).

\begin{table*}[t]
\centering
\scriptsize
\setlength{\tabcolsep}{2.2pt}
\renewcommand{\arraystretch}{0.9}
{\color{\revisioncolor}
\resizebox{\textwidth}{!}{%
\begin{tabular}{|c|
c|c|c|c|
c|c|c|c|
c|c|c|
c|c|c|}
\hline
\textbf{Load-factor} &
\multicolumn{4}{c|}{\textbf{Drop Rate} (\%)} &
\multicolumn{4}{c|}{\textbf{Queue Delay} ($\mu$s)} &
\multicolumn{3}{c|}{\textbf{Agg. Rate} (\%)} &
\multicolumn{3}{c|}{\textbf{Agg. Size} (\# pkts)} \\
$\boldsymbol{\omega} = R_{\mathrm{in}}/R_{\mathrm{out}}$ &
\textbf{Shesha} &
\textbf{FIFO} &
\textbf{iSW} &
\textbf{iSW-CA} &
\textbf{Shesha} &
\textbf{FIFO} &
\textbf{iSW} &
\textbf{iSW-CA} &
\textbf{Shesha} &
\textbf{iSW} &
\textbf{iSW-CA} &
\textbf{Shesha} &
\textbf{iSW} &
\textbf{iSW-CA} \\
\hline
1.00 & 0    & 0   & 0    & 0 & \textbf{0.29}  & 0.3   & 1.1   & 1.2   & 0.0  & 0.0  & 0.0 & 1.0  & 1.0 & 1.0 \\
\hline
1.25 & 0.1  & 19.9  & 0    & 0 & \textbf{38.0}  & 112.8  & 66.7  & 68.3  & 20.0  & 21.3 & 21.4 & 1.3  & 1.2 & 1.3 \\
\hline
1.67 & 0.5  & 39.9 & 5.8  & 0 & \textbf{87.0}  & 150.6  & 246.2 & 272.9 & 41.0  & 40.6 & 45.2 & 1.8  & 1.8 & 2.3 \\
\hline
2.50 & 10.5 & 59.9  & 18.6 & 0 & \textbf{142.6} & 225.9  & 402.3 & 412.5 & 56.0  & 57.0 & 59.8 & 2.6  & 2.4 & 2.5 \\
\hline
\end{tabular}%
}
}
\caption{{\color{\revisioncolor}Comparison of Shesha, FIFO, iSW, and iSW-CA across congestion levels. We fix the aggregation rates and agg. size of iSW and iSW-CA to be comparable to Shesha, i.e., the average number of packets compressed into one outgoing aggregate. Shesha outperforms the baselines in terms of queueing delay and/or drop rate  due to opportunistic aggregation. 
The input rate is $R_{\mathrm{in}}=100$~Gbps, and the queue depth is half the model size ($770$ packets). Aggregation rate and size are not applicable to FIFO. Details and discussion in \S\ref{sec:networkbenchmarks}.}}
\vspace{-8mm}
\label{tab:combined_output_rate_comparison}
\vspace{-1mm}
\end{table*}

\noindent\textbf{Async with periodic aggregation baselines have a drop-delay tradeoff.}
Next, we take as an example the congestion load $\omega=1.67$ and discuss the performance of the periodic aggregation baselines (similar results for different $\omega$ are in Appendix \S\ref{sec:isw_congestion_loads}). 
We set the output buffer to half the model size, i.e., $770$-packets. 
First, Tab.~\ref{tab:network_isw_isw-ca_out_varied_compact} shows that async with periodic aggregation has a clear \textit{drop-delay tradeoff}. 
With small batching windows, such as $10~\mu s$ and $50~\mu s$, it flushes segments frequently to the egress and collects a few matching segments. 
This limits aggregation and causes high packet drops at the output buffer. 
Larger batching windows improve aggregation and reduce drops, but they also increase the queueing delay.
Recall that iSW-CA avoids packet drops by keeping the flush-ready aggregation slot active when the output buffer is full which explains the high aggregation and the high queueing delay in Tab.~\ref{tab:network_isw_isw-ca_out_varied_compact}.
In comparison, Shesha outperforms in terms of queueing delay at a very low drop rate of $0.5\%$.

\begin{table*}[t]
\centering
\scriptsize
\setlength{\tabcolsep}{1.6pt}
\renewcommand{\arraystretch}{1.00}
{\color{\revisioncolor}
\begin{subtable}[t]{0.595\textwidth}
\centering
\caption{\color{\revisioncolor}Window sensitivity for iSW-style periodic aggregation.}
\label{tab:network_isw_isw-ca_out_varied_compact}
\begin{tabularx}{\linewidth}{|>{\centering\arraybackslash}p{0.12\linewidth}|*{8}{>{\centering\arraybackslash}X|}}
\hline
\multirow{2}{*}{\makecell{\textbf{Window}\\\textbf{Size}\\($\mu$s)}} &
\multicolumn{2}{c|}{\makecell{\textbf{Drop}\\\textbf{Rate} (\%)}} &
\multicolumn{2}{c|}{\makecell{\textbf{Queue}\\\textbf{Delay} ($\mu$s)}} &
\multicolumn{2}{c|}{\makecell{\textbf{Agg.}\\\textbf{Rate} (\%)}} &
\multicolumn{2}{c|}{\makecell{\textbf{Agg.} \textbf{Size}\\ (\# pkts)}} \\
\cline{2-9}
&
\textbf{iSW} &
\textbf{iSW-CA} &
\textbf{iSW} &
\textbf{iSW-CA} &
\textbf{iSW} &
\textbf{iSW-CA} &
\textbf{iSW} &
\textbf{iSW-CA} \\
\hline
10   & 36.6 & 0 & 160.5 & 258.5 & 3.2  & 39.8 & 1.05 & 1.6 \\
\hline
50   & 23.5 & 0 & 195.2 & 258.5 & 16.3 & 39.0 & 1.2  & 1.6 \\
\hline
100  & 7.1  & 0 & 242.4 & 254.5 & 32.7 & 39.7 & 1.5  & 1.5 \\
\hline
130  & 5.8  & 0 & 246.2 & 272.9 & 40.6 & 45.2 & 1.8  & 2.3 \\
\hline
500  & 0    & 0 & 317.1 & 316.8 & 73.2 & 73.7 & 3.7  & 3.8 \\
\hline
1000 & 0    & 0 & 577.2 & 575.8 & 84.5 & 84.3 & 6.4  & 6.4 \\
\hline\hline
\textbf{Shesha} &
\multicolumn{2}{c|}{\textbf{0.5}} &
\multicolumn{2}{c|}{\textbf{87.0}} &
\multicolumn{2}{c|}{\textbf{41.0}} &
\multicolumn{2}{c|}{\textbf{1.8}} \\
\hline
\end{tabularx}
\end{subtable}
\hfill
\begin{subtable}[t]{0.395\textwidth}
\centering
\caption{\color{\revisioncolor}Queue-depth sensitivity.}
\label{tab:queue_depth_delay_sensitivity_compact}
\begin{tabularx}{\linewidth}{|>{\centering\arraybackslash}p{0.24\linewidth}|*{4}{>{\centering\arraybackslash}X|}}
\hline
\multirow{2}{*}{\makecell{\textbf{Queue}\\\textbf{Depth}\\\textbf{(Packets)}}} &
\multicolumn{4}{c|}{\makecell{\textbf{Queue Delay}\\\textbf{($\mu$s)}}} \\
\cline{2-5}
&
\textbf{Shesha} &
\textbf{iSW} &
\textbf{iSW-CA} &
\textbf{FIFO} \\
\hline
770  & \textbf{87.0}  & 232.5 & 248.0 & 150.6 \\
\hline
924  & \textbf{97.4}  & 262.5 & 278.0 & 180.6 \\
\hline
1078 & \textbf{101.4} & 292.5 & 309.0 & 210.7 \\
\hline
1232 & \textbf{101.1} & 322.5 & 339.0 & 240.7 \\
\hline
1386 & \textbf{101.1} & 352.5 & 369.0 & 270.8 \\
\hline
1540 & \textbf{100.9} & 382.6 & 398.0 & 300.8 \\
\hline
\end{tabularx}
\end{subtable}
}
\caption{{\color{\revisioncolor}(a) \textbf{Async with periodic aggregation shows a drop-delay tradeoff}. Large batching windows lead to higher aggregations and less drops at the cost of large queuing delays. We include the window size ($130 \mu s$) that achieves similar aggregation rate of Shesha ($41\%$). There, Shesha outperforms async with periodic aggregation  with only $0.5\%$ packet drops and $87~\mu s$ queueing delay. Compared to our extension iSW-CA, Shesha reduces the queueing delay by 68\%. The queue depth is half the model size ($770$ packets). iSW-max has a much higher queuing delay and its results are provided in the text. (b) \textbf{Queuing delay} for Shesha, iSW-style async with periodic aggregation, iSW-CA, and FIFO for different output queue depths. The queue depth is varied from half the model size ($770$ packets) to the size of the entire model ($1540$ packets). Owing to \textit{opportunistic aggregation}, Shesha keeps the delay stable while the baselines suffer from increasing delays with increasing queue depth. Both tables are given for load factor $\omega = 1.67$.
}}
\vspace{-25pt}
\label{tab:isw_window_and_queue_depth_side_by_side}
\end{table*}

\noindent\textbf{Shesha minimizes the waiting times across queue depths.}
We increase the queue depth from half the size of the trained model ($770$ packets) to the size of the entire trained model ($1540$ packets) in Tab.~\ref{tab:queue_depth_delay_sensitivity_compact} for $\omega=1.67$ and aggregation window size $130 \mu$s. 
Tab.~\ref{tab:queue_depth_delay_sensitivity_compact} shows that the iSW-design baselines\footnote{We use the term iSW-design to refer to both methods, iSW and iSW-CA.} incur higher delays as the queue depth increases. 
Fixed window batching creates bursts of flushed aggregates that wait in the output buffer before transmission. 
A deeper output buffer absorbs more bursts, but it also keeps packets in the switch longer. 
iSW-CA avoids drops by design through holding flush-ready aggregates when the output buffer is full, however increasing the queueing delay.
SheshaQueue shows a low drop rate $(0.5\%)$ and a stable delay (at $100~\mu s$). Shesha achieves this by \textit{opportunistically aggregating} arriving packets \emph{directly} into the ones \emph{in the queue}.

}



{\color{\revisioncolor}

\subsection{Shesha improves training speed}
\label{sec:live_training}
\vspace{-2pt}

{\color{\revisioncolor}
We use a distributed PPO (Proximal policy optimization) setup with 81 workers, each running identical instances of the environment and training independently, using the RLlib library~\cite{liang2018rllib}. 
We use the environments
from Gymnasium~\cite{towers2024gymnasium}.
Upon completion of each training episode, worker~$k$ \textit{asynchronously} transmits to the PS its gradient ${g}_k$ and mean episode reward $r_k$.
The parameter server keeps the best global reward $r_{gl}$, initialized to $-\infty$. If $r_k > r_{gl}$, the parameter server updates the global weights using gradient descent as $w \leftarrow w + \gamma g_a$, where $\gamma=0.001$ is a predefined learning rate and $g_a=\texttt{avg}(g_a,g_k)$ is the average gradient~\cite{ruder2016overview}. This ensures that the global weights keep improving smoothly. 
At startup, all workers send their initial weights, and the PS initializes $w$ using the first received model. Shesha applies the normalized sum as a single update.
When multiple gradients are aggregated, the dataplane forwards the
aggregated value together with the aggregation count to the PS,
which normalizes the update before applying.
}

\begin{table*}[t]
\centering
\scriptsize
\setlength{\tabcolsep}{2.2pt}
\renewcommand{\arraystretch}{0.9}
\newlength{\liveBlockFigHeight}
\setlength{\liveBlockFigHeight}{31mm}
{\color{\revisioncolor}
\begin{subtable}[t]{0.495\textwidth}
\centering
\caption{\color{\revisioncolor}LunarLander}
\label{tab:lunar_live_block}
\begin{tabularx}{\linewidth}{|>{\centering\arraybackslash}p{0.16\linewidth}|
>{\centering\arraybackslash}X|
>{\centering\arraybackslash}X|
>{\centering\arraybackslash}X|
>{\centering\arraybackslash}X|
>{\centering\arraybackslash}X|
>{\centering\arraybackslash}X|}
\hline
\multirow{3}{*}{\textbf{Scheme}}
& \multicolumn{3}{c|}{\textbf{TTR} (minutes)}
& \multicolumn{3}{c|}{\textbf{Final Reward}} \\
\cline{2-7}
& \multicolumn{3}{c|}{$\boldsymbol{\omega}$}
& \multicolumn{3}{c|}{$\boldsymbol{\omega}$} \\
& $\mathbf{1.6}$ & $\mathbf{1.8}$ & $\mathbf{2.0}$
& $\mathbf{1.6}$ & $\mathbf{1.8}$ & $\mathbf{2.0}$ \\
\hline
FIFO      & 166.8 & 180.0 & 180.0 & 200 & 178.2 & 152.4 \\ \hline
Shesha    & \textbf{28.2} & \textbf{33.0} & \textbf{38.4} & \textbf{200} & \textbf{200} & \textbf{200} \\ \hline
iSW-CA    & 37.8 & 49.2 & 55.8 & 200 & 200 & 200 \\ \hline
iSW\_5ms  & 106.8 & 170.4 & 180.0 & 200 & 194.5 & 175.4 \\ \hline
iSW\_10ms & 90.0 & 147.6 & 180.0 & 200 & 200 & 184.1 \\ \hline
iSW\_15ms & 83.4 & 133.8 & 180.0 & 200 & 200 & 190.2 \\ \hline
iSW\_20ms & 93.0 & 180.0 & 180.0 & 200 & 199.2 & 181.6 \\ \hline
iSW\_25ms & 115.8 & 180.0 & 180.0 & 200 & 186.3 & 166.9 \\ \hline
iSW-max  & 64.8 & 65.4 & 66.0 & 200 & 200 & 200 \\ \hline
\end{tabularx}
\end{subtable}
\hfill
\begin{subtable}[t]{0.495\textwidth}
\centering
\caption{\color{\revisioncolor}HalfCheetah}
\label{tab:halfcheetah_live_block}
\begin{tabularx}{\linewidth}{|>{\centering\arraybackslash}p{0.16\linewidth}|
>{\centering\arraybackslash}X|
>{\centering\arraybackslash}X|
>{\centering\arraybackslash}X|
>{\centering\arraybackslash}X|
>{\centering\arraybackslash}X|
>{\centering\arraybackslash}X|}
\hline
\multirow{3}{*}{\textbf{Scheme}}
& \multicolumn{3}{c|}{\textbf{TTR} (minutes)}
& \multicolumn{3}{c|}{\textbf{Final Reward}} \\
\cline{2-7}
& \multicolumn{3}{c|}{$\boldsymbol{\omega}$}
& \multicolumn{3}{c|}{$\boldsymbol{\omega}$} \\
& $\mathbf{1.6}$ & $\mathbf{1.8}$ & $\mathbf{2.0}$
& $\mathbf{1.6}$ & $\mathbf{1.8}$ & $\mathbf{2.0}$ \\
\hline
FIFO      & 180.0 & 180.0 & 180.0 & 2471.9 & 2194.9 & 1954.6 \\ \hline
Shesha    & \textbf{54.6} & \textbf{63.0} & \textbf{73.2} & \textbf{3086.7} & \textbf{3090.1} & {3029.1} \\ \hline
iSW-CA    & 75.0 & 91.8 & 100.8 & 3080.1 & 3027.8 & 3033.8 \\ \hline
iSW\_5ms  & 172.8 & 180.0 & 180.0 & 2948.4 & 2843.0 & 2620.0 \\ \hline
iSW\_10ms & 97.8 & 177.0 & 180.0 & 3027.1 & 3030.6 & 2725.1 \\ \hline
iSW\_15ms & 94.2 & 171.0 & 180.0 & 3038.2 & 3032.7 & 2769.3 \\ \hline
iSW\_20ms & 99.6 & 178.8 & 180.0 & 3019.1 & 3007.6 & 2929.2 \\ \hline
iSW\_25ms & 109.2 & 180.0 & 180.0 & 2920.0 & 2854.1 & 2812.3 \\ \hline
iSW-max  & 119.8 & 120.2 & 118.5 & 3037.2 & 3033.5 & \textbf{3048.1} \\ \hline
\end{tabularx}
\end{subtable}
}
\vspace{0.5mm}

{\color{\revisioncolor}
\begin{subtable}[t]{0.425\textwidth}
\centering
\caption{\color{\revisioncolor}Atari Pong}
\label{tab:pong_live_block}
\begin{tabularx}{\linewidth}{|>{\centering\arraybackslash}p{0.19\linewidth}|
>{\centering\arraybackslash}X|
>{\centering\arraybackslash}X|
>{\centering\arraybackslash}X|
>{\centering\arraybackslash}X|
>{\centering\arraybackslash}X|
>{\centering\arraybackslash}X|}
\hline
\multirow{3}{*}{\textbf{Scheme}}
& \multicolumn{3}{c|}{\textbf{TTR} (minutes)}
& \multicolumn{3}{c|}{\textbf{Final Reward}} \\
\cline{2-7}
& \multicolumn{3}{c|}{$\boldsymbol{\omega}$}
& \multicolumn{3}{c|}{$\boldsymbol{\omega}$} \\
& $\mathbf{1.6}$ & $\mathbf{1.8}$ & $\mathbf{2.0}$
& $\mathbf{1.6}$ & $\mathbf{1.8}$ & $\mathbf{2.0}$ \\
\hline
FIFO      & 504.0 & 504.0 & 504.0 & 6.1 & -1.2 & -7.6 \\ \hline
Shesha    & \textbf{85.8} & \textbf{147.0} & \textbf{214.8} & \textbf{18.4} & \textbf{18.9} & \textbf{18.6} \\ \hline
iSW-CA    & 105.6 & 178.8 & 260.4 & 18.3 & 18.1 & 17.8 \\ \hline
iSW\_10ms & 378 & 435.2 & 532.1 & 12.8 & 11.3 & 10.6 \\ \hline
iSW\_25ms & 224.4 & 319.2 & 507.6 & 16.5 & 17.4 & 15.6 \\ \hline
iSW\_35ms & 148.2 & 276.0 & 433.8 & 18.0 & 18.3 & 16.9 \\ \hline
iSW\_45ms & 170.4 & 243.6 & 415.2 & 17.3 & 17.0 & 17.2 \\ \hline
iSW\_55ms & 228.0 & 291.0 & 402.0 & 16.3 & 16.6 & 17.1 \\ \hline
iSW-max  & 245.2 & 242.9 & 253.5 & 18.3  & 18.2 & 18.0 \\ \hline
\end{tabularx}
\end{subtable}
\hfill
\begin{subfigure}[t]{0.565\textwidth}
\centering
\caption{\color{\revisioncolor}Summary of absolute TTR for varying congestion.}
\label{fig:additional_time_common_reward_isw_iswca}
\vspace{1pt}
\includegraphics[height=\liveBlockFigHeight,width=\linewidth,keepaspectratio]
{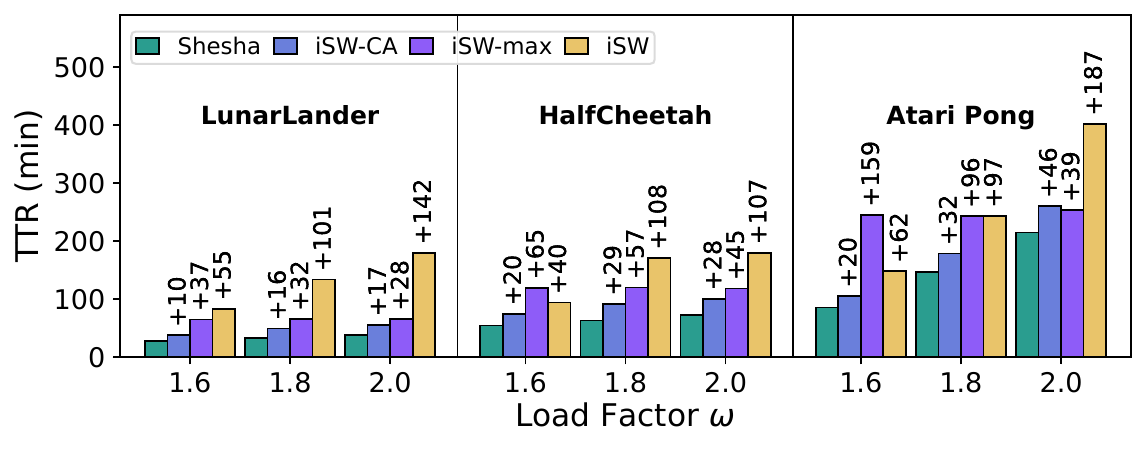}

\end{subfigure}
}
\caption{{\color{\revisioncolor}
\textbf{Shesha is faster in live-training across three diverse DRL workloads.}
Time-to-Target Reward (TTR) is given in minutes. 
(a)-(c) Shesha achieves the smallest TTR for all configurations and congestion load factors $\omega$. (d) Absolute time required by Shesha, iSW-CA, iSW, and iSW-max to achieve the final reward threshold. Numbers above the bars show the additional TTR in minutes relative to Shesha. For iSW, we use the best-performing fixed window configuration for each workload and load factor. FIFO performs poorly as depicted in Fig.~\ref{fig:fifo_TTR_penalty} in the appendix.}
}
\vspace{-28pt}
\label{tab:live_training_summary_blocks}
\end{table*}

For live training, we report the \textbf{Time-to-Target Reward} (TTR)\footnote{Similar to Time-to-Accuracy (TTA) in supervised learning settings.} as well as the final reward achieved in training until the standard training horizon. 
TTR is the wall-clock time needed to reach the workload-specific reward target. 
We report the final reward as the average worker reward at the end of training. 
If a scheme does not reach the target within the training horizon, we report the horizon time (cf. Tab.~\ref{tab:workload_characterization_all}) and the final reward observed at that point.

Our live training testbed consists of $81$ physical workers, a Tofino switch, an Alveo U55C, and a parameter server  as described at the beginning of this section. 
Each worker trains locally and sends segmented gradient updates after an iteration, creating time-varying congestion for the in-network aggregation. 
Workers do not preempt the current iteration when a newer global model arrives and apply the received model at the next iteration boundary. 

\noindent\textbf{Shesha accelerates Training across diverse Workloads.} 
We evaluate three DRL workloads with different characteristics, i.e., low-dimensional discrete control in \textbf{\texttt{LunarLander-v3}}, continuous control in \textbf{\texttt{HalfCheetah-v4}}, and image-based training in \textbf{\texttt{Pong}} (Atari 2600). Details of these workloads are in Tab.~\ref{tab:workload_characterization_all} in the appendix. Each workload uses segmented model updates, and we set the queue depth of the baseline to half the number of packets of one full model update ($770$ packets). 
We use large window sizes for \textbf{\texttt{Pong}} as we observe low final reward for window sizes $<25$ms (cf. Tab.~\ref{tab:pong_live_block}). 

Tab.~\ref{tab:halfcheetah_live_block}-\ref{tab:pong_live_block} report the \textbf{Time-to-Target Reward} and the final reward at the end of training for different congestion load factors $\omega$.
Tab.~\ref{tab:live_training_summary_blocks} shows that \textit{Shesha reaches the target reward faster than all the baselines across all configurations and all three workloads}. 
Again, Shesha outperforms iSW by avoiding fixed-window batching. Our live training results reaffirm that the iSW performance is sensitive to tuning the batching window, depending on the combination of workload and congestion load. That is, small batching windows limit the aggregation rate and lead to drops under congestion, while large batching windows delay the updates. Shesha does not hold updates and, essentially, through opportunistic aggregation, handles the potential contention that arises from not holding updates.  Recall that iSW-max maximizes the aggregation rate. However, its TTR suffers due to the high delay until the aggregate is distributed.  Closest to Shesha is iSW-CA\footnote{\color{\revisioncolor}All evaluated iSW-CA window sizes show similar TTR and final reward, so we report one representative configuration.} as it avoids dropping flush-ready aggregates when the output queue is full but it is still based on the batching window aggregation that holds updates. Workers continue training with older model versions during this waiting time, which increases staleness. 
FIFO performs the worst as it does not aggregate updates and loses packets under congestion.
Fig.~\ref{fig:additional_time_common_reward_isw_iswca} summarizes the absolute TTR for Shesha, iSW-CA, iSW-max, and iSW, where for iSW we use \textit{the best possible fixed-window configuration in each setting}. Shesha is consistently faster across all workloads. Fig.~\ref{fig:normalized_ttr_summary}  
in the appendix shows the speedup reaching up to  $4.7\times$ for best iSW, $1.5\times$ for iSW-CA, $2.9\times$ for iSW-max, and $8.5\times$ for FIFO. 

{\color{\revisioncolor}

\subsection{Multi-tier Topology}
\label{sec:multihop_ns3}

We use ns-3 simulations~\cite{ns3} to evaluate Shesha in a larger
multi-tier topology reporting \textit{AoM degradation}. 
It measures the percentage increase in AoM when incrementing the number of training clusters in the network from one to $N$.
Recall from \S\ref{subsec:AoM} that a larger AoM means a more stale global model. 
Fig.~\ref{fig:aom_deg_s2} shows the simulated topology consisting of two access switches, $SW_1$ and $SW_2$,
and one \textit{shared bottleneck }upstream switch, $SW_3$, connected to the PS. 
Each access switch serves five clusters where each cluster has ten workers.
We group the clusters behind $SW_1$ as $S1$ and the clusters behind
$SW_2$ as $S2$. 
We compare FIFO with two variants of Shesha: Shesha without any worker-side logic, and Shesha\_TC, which incorporates the worker-side \textit{transmission control} algorithm from §~\ref{sec:worker_algo}. 
Shesha aggregates only packets from the same cluster
and the same model segment. It never aggregates packets across clusters or across $S1$ and $S2$. 
%
We focus here on evaluating Shesha's AoM which is used to measure model freshness.
For each cluster, we first run a \textit{standalone} experiment where only that cluster sends updates and denote the result as $AoM_{std}$. 
We then run a \textit{shared} experiment where both $S1$ and $S2$ send updates, and denote the result as $AoM_{shared}$. 
We compute AoM degradation as the percentage increase from $AoM_{std}$ to $AoM_{shared}$, i.e., the impact of sharing the network on the model staleness. 

\noindent\textbf{Homogeneous sharing.}
Tab.~\ref{tab:aom_degradation_combined} shows the homogeneous case, where both clusters generate updates at the same rate. 
FIFO increases the AoM of both clusters by more than $250\%$. 
Since the network conditions and transmission rates are uniform across all clusters, both Shesha variants perform identically. 
Shesha reduces the AoM degradation to 
$\leq10\%$ for both clusters showing that opportunistic aggregation reduces the freshness penalty from sharing the bottleneck.





\begin{table*}[t]
\centering
\scriptsize
\setlength{\tabcolsep}{1pt}
\captionsetup{skip=2pt}
\renewcommand{\arraystretch}{0.98}

\begin{subfigure}[t]{0.4\textwidth}
    \centering
    \caption{Multi-tier topology with 10 clusters.}
    \includegraphics[width=\linewidth]{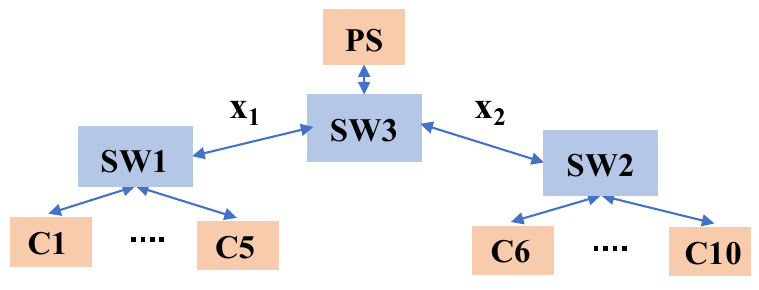}
    \label{fig:aom_deg_s2}
\end{subfigure}
\hfill
{\color{\revisioncolor}
\begin{subtable}[t]{0.59\textwidth}
\centering
\caption{\color{\revisioncolor}AoM degradation under homogeneous and heterogeneous workloads. AoM values are in msec.}
\label{tab:aom_degradation_combined}
\resizebox{\linewidth}{!}{%
\begin{tabular}{|l|c|c|c|c|c|c|c|}
\hline
\makecell{\textbf{Queue}\\\textbf{Config}} &
\makecell{$AoM_{\mathrm{std}}$\\S1} &
\makecell{$AoM_{\mathrm{shared}}$\\S1} &
\makecell{S1\\Deg. (\%)} &
\makecell{$AoM_{\mathrm{std}}$\\S2} &
\makecell{$AoM_{\mathrm{shared}}$\\S2} &
\makecell{S2\\Deg. (\%)} &
\makecell{Deg.\\Gap} \\
\hline

\multicolumn{8}{|l|}{\textbf{Homogeneous workload}} \\
\hline
FIFO   & 60 & 214 & 256.7 & 60  & 212 & 253.3 & 3.4 \\
\hline
Shesha & 60 & 66  & 10.0  & 60  & 65  & 8.3   & 1.7 \\
\hline

\multicolumn{8}{|l|}{\textbf{Heterogeneous workload}} \\
\hline
FIFO       & 60 & 88 & 46.7 & 160 & 356 & 122.5 & 75.8 \\
\hline
Shesha     & 60 & 65 & 8.3  & 160 & 184 & 15.0  & 6.7 \\
\hline
Shesha\_TC & 60 & 68 & 13.3 & 160 & 176 & 10.0  & 3.3 \\
\hline

\end{tabular}%
}
\end{subtable}
}
\caption{\color{\revisioncolor}Topology and AoM degradation results for homogeneous and heterogeneous workloads.}
\vspace{-5mm}
\label{tabfig:aom_topology_and_degradation}
\end{table*}

\noindent\textbf{Heterogeneous sharing.}
Tab.~\ref{tab:aom_degradation_combined} shows the heterogeneous
case, where $S1$ and $S2$ generate updates at different rates. 
The degradation metric shows the impact of sharing, i.e.,
FIFO increases AoM degradation to $46.7\%$ for $S1$ and $122.5\%$ for $S2$ with a gap of $75.8\%$.
Shesha significantly lowers the degradation gap to $6.7$\%. 
Shesha\_TC further reduces the degradation gap by adapting worker transmissions using queue feedback. It balances the freshness penalty across heterogeneous clusters. We further study AoM degradation under asymmetric link capacities in Fig.~\ref{fig:aom_deg_s1} in Appendix \S\ref{sec:isw_congestion_loads}.
}
}


\section{Related Work}
\label{sec:relatedwork}
Prior works on in-network aggregation for DML primarily rely on programmable switches, such as Intel Tofino~\cite{christianos2021scaling,lao2021atp,sapio2021switchML} 
or on-chip FPGAs~\cite{bressana2020trading,iSW, itsubo2020accelerating, ma2022fpga}.
Designing synchronous arrivals of all updates as in SwitchML~\cite{sapio2021switchML} enables low-latency in-network aggregation, which makes it, however, not suited for heterogeneous environments with stragglers or, in general, for asynchronous training. Additionally, it relies on static aggregator allocation per job, limiting resource flexibility in multi-tenant settings.
ATP~\cite{lao2021atp} improves on this by supporting multi-level, best-effort aggregation and dynamic resource sharing. However, it still suffers from stragglers due to the periodic, synchronized aggregation rounds. In addition, it has limited scalability due to its reliance on bitmap-based resource allocation and the restricted memory model of Tofino. 

Beyond Tofino-based systems, there exists a number of FPGA-based solutions~\cite{bressana2020trading,iSW, itsubo2020accelerating, ma2022fpga}.
Trio~\cite{yang2022trio} introduces a custom chipset to enable in-network aggregation and straggler mitigation. However, its timer-thread-based straggler handling can become slower under extremely large-scale workloads.
SHArP~\cite{graham2016sharp} targets in-network aggregation for collective operations (e.g., all-reduce) using static reduction trees implemented in network switches, which lack support for dynamic or asynchronous aggregation, and cannot adapt to multi-tenant training workloads or per-update decision logic like AoM.
Finally, iSW~\cite{iSW} proposes an asynchronous RL framework, but the asynchrony arises only from non-blocking model reception. The in-network aggregation remains synchronous, and like SwitchML and ATP, iSW stores aggregated gradients on the switch, impacting scalability. Further, supporting large-scale or multi-tenant jobs in iSW requires manual buffer scaling, adding system complexity.

In contrast, Shesha's opportunistic aggregation and the use of on-chip FPGA memory to store updates in the queue, hence, quickly processing those updates at line rate. As Shesha only keeps aggregated gradients while they are in the queue, as opposed to retaining them in the switch registers, it avoids state explosion and scales efficiently across many jobs and workers, without being limited by fixed switch resources. Importantly, Shesha’s design decouples end-host and network logic. While a worker-side transmission control algorithm can decrease AoM degradation, Shesha’s acceleration engine alone is sufficient to support AoM-reducing opportunistic aggregation without requiring any modifications at the end host.
\section{Conclusion}
\label{sec:conslusion}
We present Shesha, a programmable in-network accelerator tailored for asynchronous  Distributed Reinforcement Learning. 
Shesha addresses model staleness caused due to congestion by introducing opportunistic on the fly model update aggregation on the data plane. 
To this end, we introduce a particular queue design, \texttt{SheshaQueue}, which enables opportunistic aggregation and replacement.  
Implemented on a hybrid P$4$-FPGA architecture, Shesha achieves significant improvements in model update timeliness and system throughput, reducing update losses and accelerating training convergence. 
We provide a worker-side transmission control to handle congestion as well as a formal quantification of the model staleness, denoted Age-of-Model, that allows verifying formalized operation objectives over multiple concurrent clusters.

\begin{acks}
This work was supported in part by the ENVELOPE project under the SNS JU (Grant No. 101139048) and the L3S Research Center, Hannover, Germany.
\end{acks}

\balance
\enlargethispage{0pt}
\bibliographystyle{plainurl}
\balance
\bibliography{OLAF-arxiv/reference}

@inproceedings{lao2021atp,
  title={{ATP: In-network aggregation for multi-tenant learning}},
  author={Lao, ChonLam and Le, Yanfang and Mahajan, Kshiteej and Chen, Yixi and Wu, Wenfei and Akella, Aditya and Swift, Michael},
  booktitle={Proceedings of the 18th USENIX Symposium on Networked Systems Design and Implementation},
  pages={741--761},
  year={2021},
  url={https://www.usenix.org/conference/nsdi21/presentation/lao}
}

@article{ruder2016overview,
  title={{An overview of gradient descent optimization algorithms}},
  author={Ruder, Sebastian},
  journal={arXiv preprint arXiv:1609.04747},
  year={2016},
  doi={https://doi.org/10.48550/arXiv.1609.04747}
}

@article{towers2024gymnasium,
  title={Gymnasium: a standard interface for reinforcement learning environments},
  author={Kwiatkowski, Ariel and Towers, Mark and Terry, JK and Balis, John U and De Cola, Gianluca and Deleu, Tristan and Goul{\~a}o, Manuel and Andreas, Kallinteris and Krimmel, Markus and KG, Arjun and others},
  year={2024},
  url={https://openreview.net/forum?id=feFlfuOse1}
}

@inproceedings{liang2018rllib,
  title={{RLlib: Abstractions for distributed reinforcement learning}},
  author={Liang, Eric and Liaw, Richard and Nishihara, Robert and Moritz, Philipp and Fox, Roy and Goldberg, Ken and Gonzalez, Joseph and Jordan, Michael and Stoica, Ion},
  booktitle={Proceedings of the International conference on machine learning},
  pages={3053--3062},
  year={2018},
  url={https://proceedings.mlr.press/v80/liang18b}
}

@article{li2014communication,
  title={Communication efficient distributed machine learning with the parameter server},
  author={Li, Mu and Andersen, David G and Smola, Alexander and Yu, Kai},
  journal={Advances in neural information processing systems},
  volume={27},
  year={2014},
  url={https://papers.nips.cc/paper_files/paper/2014/hash/935ad074f32d1e8f085a143449894cdc-Abstract.html}
}

@inproceedings{sapio2021switchML,
  title={{Scaling distributed machine learning with In-Network aggregation}},
  author={Sapio, Amedeo and Canini, Marco and Ho, Chen-Yu and Nelson, Jacob and Kalnis, Panos and Kim, Changhoon and Krishnamurthy, Arvind and Moshref, Masoud and Ports, Dan and Richt{\'a}rik, Peter},
  booktitle={Proceedings of the 18th USENIX Symposium on Networked Systems Design and Implementation},
  pages={785--808},
  year={2021},
  url={https://www.usenix.org/conference/nsdi21/presentation/sapio}
}

@inproceedings{iSW,
  title={{Accelerating distributed reinforcement learning with in-switch computing}},
  author={Li, Youjie and Liu, Iou-Jen and Yuan, Yifan and Chen, Deming and Schwing, Alexander and Huang, Jian},
  booktitle={Proceedings of the 46th International Symposium on Computer Architecture},
  pages={279--291},
  year={2019},
  doi={https://doi.org/10.1145/3307650.3322259}
}

@inproceedings{yang2022trio,
  title={{Using Trio: Juniper Networks' programmable chipset-for emerging in-network applications}},
  author={Yang, Mingran and Baban, Alex and Kugel, Valery and Libby, Jeff and Mackie, Scott and Kananda, Swamy Sadashivaiah Renu and Wu, Chang-Hong and Ghobadi, Manya},
  booktitle={Proceedings of the ACM SIGCOMM Conference},
  pages={633--648},
  year={2022},
  doi={https://doi.org/10.1145/3544216.3544262}
}

@article{schulman2017proximal,
  title={{Proximal policy optimization algorithms}},
  author={Schulman, John and Wolski, Filip and Dhariwal, Prafulla and Radford, Alec and Klimov, Oleg},
  journal={arXiv preprint arXiv:1707.06347},
  year={2017},
  doi={https://doi.org/10.48550/arXiv.1707.06347}
}

@article{rafailov2023direct,
  title={{Direct preference optimization: Your language model is secretly a reward model}},
  author={Rafailov, Rafael and Sharma, Archit and Mitchell, Eric and Manning, Christopher D and Ermon, Stefano and Finn, Chelsea},
  journal={Advances in Neural Information Processing Systems},
  volume={36},
  pages={53728--53741},
  year={2023},
  url={https://proceedings.neurips.cc/paper_files/paper/2023/hash/a85b405ed65c6477a4fe8302b5e06ce7-Abstract-Conference.html}
}

@inproceedings{doshi2021distributedPPO,
  title={{Distributed proximal policy optimization for contention-based spectrum access}},
  author={Doshi, Akash and Andrews, Jeffrey G},
  booktitle={Proceedings of the 55th Asilomar Conference on Signals, Systems, and Computers},
  pages={340--344},
  year={2021},
  doi={https://doi.org/10.1109/IEEECONF53345.2021.9723270}
}

@misc{amdU55C,
  author       = {{AMD}},
  title        = {{AMD Alveo U55C Data Center Accelerator Card}},
  howpublished          = {\url{https://www.amd.com/en/products/accelerators/alveo/u55c/a-u55c-p00g-pq-g.html}},
  note         = {Accessed: 2025-04-04},
year={2025}
}

@misc{intelTofino,
  author       = {{Intel Corporation}},
  title        = {{Intel{\textregistered} Tofino Intelligent Fabric Processors}},
  howpublished          = {\url{https://www.intel.com/content/www/us/en/products/details/network-io/intelligent-fabric-processors/tofino.html}},
  note         = {Accessed: 2025-04-04},
year={2025}
}

@misc{ns3,
  author       = {{ns-3 Project}},
  title        = {{ns-3 Network Simulator}},
  howpublished          = {\url{https://www.nsnam.org/}},
  note         = {Accessed: 2025-04-04}
}

@inproceedings{christianos2021scaling,
  title={{Scaling multi-agent reinforcement learning with selective parameter sharing}},
  author={Christianos, Filippos and Papoudakis, Georgios and Rahman, Muhammad A and Albrecht, Stefano V},
  booktitle={Proceedings of the International Conference on Machine Learning},
  pages={1989--1998},
  year={2021},
  url={https://proceedings.mlr.press/v139/christianos21a.html}
}

@inproceedings{cacc_mina,
  title={{Toward formally verifying congestion control behavior}},
  author={Arun, Venkat and Arashloo, Mina Tahmasbi and Saeed, Ahmed and Alizadeh, Mohammad and Balakrishnan, Hari},
  booktitle={Proceedings of the ACM SIGCOMM Conference},
  pages={1--16},
  year={2021},
  DOI={https://doi.org/10.1145/3452296.3472912}
}

@inproceedings{graham2016sharp,
  title={{Scalable hierarchical aggregation protocol (SHArP): A hardware architecture for efficient data reduction}},
  author={Graham, Richard L and Bureddy, Devendar and Lui, Pak and Rosenstock, Hal and Shainer, Gilad and Bloch, Gil and Goldenerg, Dror and Dubman, Mike and Kotchubievsky, Sasha and Koushnir, Vladimir and others},
  booktitle={Proceedings of the First International Workshop on Communication Optimizations in HPC (COMHPC)},
  pages={1--10},
  year={2016},
  doi={https://doi.org/10.1109/COMHPC.2016.006}
}

@inproceedings{bressana2020trading,
  title={{Trading latency for compute in the network}},
  author={Bressana, Pietro and Zilberman, Noa and Vucinic, Dejan and Soul{\'e}, Robert},
  booktitle={Proceedings of the Workshop on Network Application Integration/CoDesign},
  pages={35--40},
  year={2020},
  doi={https://doi.org/10.1145/3405672.3405807}
}

@inproceedings{itsubo2020accelerating,
  title={{Accelerating deep learning using multiple GPUs and FPGA-based 10GbE switch}},
  author={Itsubo, Tomoya and Koibuchi, Michihiro and Amano, Hideharu and Matsutani, Hiroki},
  booktitle={Proceedings of the 28th Euromicro International Conference on Parallel, Distributed and Network-Based Processing (PDP)},
  pages={102--109},
  year={2020},
  doi={https://doi.org/10.1109/PDP50117.2020.00022}
}

@article{ma2022fpga,
  title={{FPGA-based AI smart NICs for scalable distributed AI training systems}},
  author={Ma, Rui and Georganas, Evangelos and Heinecke, Alexander and Gribok, Sergey and Boutros, Andrew and Nurvitadhi, Eriko},
  journal={IEEE Computer Architecture Letters},
  volume={21},
  number={2},
  pages={49--52},
  year={2022},
  doi={https://doi.org/10.1109/LCA.2022.3189207}
}

@misc{xilinxOpenNIC,
  author       = {{Xilinx}},
  title        = {{Open-NIC Project}},
  howpublished          = {\url{https://github.com/Xilinx/open-nic}},
  note         = {Accessed: 2025-04-04},
year={2025}
}

@misc{amdVitisNetP4,
  author       = {{AMD}},
  title        = {{VitisNetP4 IP for Adaptive SoCs and FPGAs}},
  howpublished          = {\url{https://www.amd.com/en/products/adaptive-socs-and-fpgas/intellectual-property/ef-di-vitisnetp4.html}},
  note         = {Accessed: 2025-04-04},
  year={2025}
}

@article{bosshart2014p4,
  title={{P4: Programming protocol-independent packet processors}},
  author={Bosshart, Pat and Daly, Dan and Gibb, Glen and Izzard, Martin and McKeown, Nick and Rexford, Jennifer and Schlesinger, Cole and Talayco, Dan and Vahdat, Amin and Varghese, George and others},
  journal={ACM SIGCOMM Computer Communication Review},
  volume={44},
  number={3},
  pages={87--95},
  year={2014},
  doi={https://doi.org/10.1145/2656877.2656890}
}

@misc{axi4stream,
  author       = {{ARM Ltd.}},
  title        = {{AMBA AXI4-Stream Protocol Specification}},
  year         = {2010},
  howpublished = {\url{https://developer.arm.com/documentation/ihi0051}},
  note         = {Accessed: 2025-04-04}
}

@article{liu2020high,
  title={{High-throughput synchronous deep RL}},
  author={Liu, Iou-Jen and Yeh, Raymond and Schwing, Alexander},
  journal={Advances in Neural Information Processing Systems},
  volume={33},
  pages={17070--17080},
  year={2020},
  url={https://proceedings.neurips.cc/paper/2020/hash/c6447300d99fdbf4f3f7966295b8b5be-Abstract.html}
}

@article{das2016distributed,
  title={{Distributed deep learning using synchronous stochastic gradient descent}},
  author={Das, Dipankar and Avancha, Sasikanth and Mudigere, Dheevatsa and Vaidynathan, Karthikeyan and Sridharan, Srinivas and Kalamkar, Dhiraj and Kaul, Bharat and Dubey, Pradeep},
  journal={arXiv preprint arXiv:1602.06709},
  year={2016}
}

@article{verbraeken2020survey,
  title={{A survey on distributed machine learning}},
  author={Verbraeken, Joost and Wolting, Matthijs and Katzy, Jonathan and Kloppenburg, Jeroen and Verbelen, Tim and Rellermeyer, Jan S},
  journal={ACM Computing Surveys (csur)},
  volume={53},
  number={2},
  pages={1--33},
  year={2020}
}

@inproceedings{li2013parameter,
  title={{Parameter server for distributed machine learning}},
  author={Li, Mu and Zhou, Li and Yang, Zichao and Li, Aaron and Xia, Fei and Andersen, David G and Smola, Alexander},
  booktitle={Proceedings of the Big learning NIPS workshop},
  volume={6},
  number={2},
  year={2013},
  url={https://www.istc-cc.cmu.edu/publications/papers/2013/ps.pdf}
}

@inproceedings{li2024ps,
author = {Li, Mu and Andersen, David G. and Park, Jun Woo and Smola, Alexander J. and Ahmed, Amr and Josifovski, Vanja and Long, James and Shekita, Eugene J. and Su, Bor-Yiing},
title = {{Scaling distributed machine learning with the parameter server}},
year = {2014},
booktitle = {Proceedings of the 11th USENIX Conference on Operating Systems Design and Implementation},
url={https://www.usenix.org/system/files/conference/osdi14/osdi14-paper-li_mu.pdf}
}

@article{dean2012large,
  title={{Large scale distributed deep networks}},
  author={Dean, Jeffrey and Corrado, Greg and Monga, Rajat and Chen, Kai and Devin, Matthieu and Mao, Mark and Ranzato, Marc'aurelio and Senior, Andrew and Tucker, Paul and Yang, Ke and others},
  journal={Advances in neural information processing systems},
  volume={25},
  year={2012},
  url={https://proceedings.neurips.cc/paper_files/paper/2012/hash/6aca97005c68f1206823815f66102863-Abstract.html}
}

@article{liu2024asynchronous,
  title={{Asynchronous local-sgd training for language modeling}},
  author={Liu, Bo and Chhaparia, Rachita and Douillard, Arthur and Kale, Satyen and Rusu, Andrei A and Shen, Jiajun and Szlam, Arthur and Ranzato, Marc'Aurelio},
  journal={arXiv preprint arXiv:2401.09135},
  year={2024},
  doi={https://doi.org/10.48550/arXiv.2401.09135}
}

@article{wijmans2019dd,
  title={{DD-PPO: Learning near-perfect pointgoal navigators from 2.5 billion frames}},
  author={Wijmans, Erik and Kadian, Abhishek and Morcos, Ari and Lee, Stefan and Essa, Irfan and Parikh, Devi and Savva, Manolis and Batra, Dhruv},
  journal={arXiv preprint arXiv:1911.00357},
  year={2019},
  doi={https://doi.org/10.48550/arXiv.1911.00357}
}

@article{wang2020domain,
  title={{Domain-specific communication optimization for distributed DNN training}},
  author={Wang, Hao and Chen, Jingrong and Wan, Xinchen and Tian, Han and Xia, Jiacheng and Zeng, Gaoxiong and Wang, Weiyan and Chen, Kai and Bai, Wei and Jiang, Junchen},
  journal={arXiv preprint arXiv:2008.08445},
  year={2020},
  doi={https://doi.org/10.48550/arXiv.2008.08445}
}

@inproceedings{chen2023boosting,
  title={{Boosting distributed machine learning training through loss-tolerant transmission protocol}},
  author={Chen, Zixuan and Shi, Lei and Liu, Xuandong and Ai, Xin and Liu, Sen and Xu, Yang},
  booktitle={Proceedings of the IEEE/ACM 31st International Symposium on Quality of Service (IWQoS)},
  pages={1--10},
  year={2023},
  doi={https://doi.org/10.1109/IWQoS57198.2023.10188699}
}

@article{xiao2022asynchronous,
  title={{Asynchronous actor-critic for multi-agent reinforcement learning}},
  author={Xiao, Yuchen and Tan, Weihao and Amato, Christopher},
  journal={Advances in Neural Information Processing Systems},
  volume={35},
  pages={4385--4400},
  year={2022},
  doi={https://proceedings.neurips.cc/paper_files/paper/2022/hash/1c153788756d35559c22d105d1182c30-Abstract-Conference.html}
}

@article{amari1993backpropagation,
  title={{Backpropagation and stochastic gradient descent method}},
  author={Amari, Shun-ichi},
  journal={Neurocomputing},
  volume={5},
  number={4-5},
  pages={185--196},
  year={1993},
  doi={10.1016/0925-2312(93)90006-O},
  url={https://doi.org/10.1016/0925-2312(93)90006-O}
}

@article{yates2021age,
  title={{Age of information: An introduction and survey}},
  author={Yates, Roy D and Sun, Yin and Brown, D Richard and Kaul, Sanjit K and Modiano, Eytan and Ulukus, Sennur},
  journal={IEEE Journal on Selected Areas in Communications},
  volume={39},
  number={5},
  pages={1183--1210},
  year={2021},
  doi={https://doi.org/10.1109/JSAC.2021.3065072}
}

@article{kaelbling1996reinforcement,
  title={{Reinforcement learning: A survey}},
  author={Kaelbling, Leslie Pack and Littman, Michael L and Moore, Andrew W},
  journal={Journal of artificial intelligence research},
  volume={4},
  pages={237--285},
  year={1996},
  doi={https://doi.org/10.1613/jair.301}
}

@article{li2017deep,
  title={{Deep reinforcement learning: An overview}},
  author={Li, Yuxi},
  journal={arXiv preprint arXiv:1701.07274},
  year={2017},
  doi={https://doi.org/10.48550/arXiv.1701.07274}
}

@article{kober2013reinforcement,
  title={{Reinforcement learning in robotics: A survey}},
  author={Kober, Jens and Bagnell, J Andrew and Peters, Jan},
  journal={The International Journal of Robotics Research},
  volume={32},
  number={11},
  pages={1238--1274},
  year={2013},
  doi={https://doi.org/10.1177/0278364913495721}
}

@article{xiang2024reinforcement,
  title={{Reinforcement learning in autonomous driving}},
  author={Xiang, Dantong},
  journal={Applied and Computational Engineering},
  volume={48},
  pages={17--23},
  year={2024},
  doi={https://doi.org/10.54254/2755-2721/48/20241072}
}

@inproceedings{chen2024rina,
  title={{Rina: Enhancing Ring-AllReduce with In-network Aggregation in Distributed Model Training}},
  author={Chen, Zixuan and Liu, Xuandong and Li, Minglin and Hu, Yinfan and Mei, Hao and Xing, Huifeng and Wang, Hao and Shi, Wanxin and Liu, Sen and Xu, Yang},
  booktitle={Proceedings of the IEEE 32nd International Conference on Network Protocols (ICNP)},
  pages={1--12},
  year={2024},
  doi={https://doi.org/10.1109/ICNP61940.2024.10858570}
}

@inproceedings{wan2020rat,
  title={{Rat-resilient allreduce tree for distributed machine learning}},
  author={Wan, Xinchen and Zhang, Hong and Wang, Hao and Hu, Shuihai and Zhang, Junxue and Chen, Kai},
  booktitle={Proceedings of the 4th Asia-Pacific Workshop on Networking},
  pages={52--57},
  year={2020},
  doi={https://doi.org/10.1145/3411029.3411037}
}

@inproceedings{baganal2023rpm,
  title={{RPM: Reverse Path Congestion Marking on P4 Programmable Switches}},
  author={Baganal-Krishna, Nehal and Tran, Tuan-Dat and Kundel, Ralf and Rizk, Amr},
  booktitle={Proceedings of the IEEE 48th Conference on Local Computer Networks (LCN)},
  pages={1--4},
  year={2023},
  doi={https://doi.org/10.1109/LCN58197.2023.10223327}
}

@inproceedings{feldmann2019p4,
  title={{P4-enabled network-assisted congestion feedback: A case for nacks}},
  author={Feldmann, Anja and Chandrasekaran, Balakrishnan and Fathalli, Seifeddine and Weyulu, Emilia N},
  booktitle={Proceedings of Workshop on Buffer Sizing},
  pages={1--7},
  year={2019},
  doi={https://doi.org/10.1145/3375235.3375238}
}

@inproceedings{z3,
  title={Z3: An efficient SMT solver},
  author={De Moura, Leonardo and Bj{\o}rner, Nikolaj},
  booktitle={Proceedings of the International conference on Tools and Algorithms for the Construction and Analysis of Systems},
  pages={337--340},
  year={2008},
  doi={https://doi.org/10.1007/978-3-540-78800-3_24}
}

@misc{cisco8100,
  author       = {{Cisco Systems, Inc.}},
  title        = {{Cisco N9300 Series Smart Switches}},
  howpublished = {\url{https://www.cisco.com/c/de_de/support/switches/9300-series-smart-switches/series.html}},
  note         = {Accessed: 2025-04-04}
}

@inproceedings{mujoco,
author = {Todorov, Emanuel and Erez, Tom and Tassa, Yuval},
year = {2012},
month = {10},
pages = {5026-5033},
title = {MuJoCo: A physics engine for model-based control},
isbn = {978-1-4673-1737-5},
journal = {Proceedings of the IEEE/RSJ International Conference on Intelligent Robots and Systems},
doi = {10.1109/IROS.2012.6386109}
}

@article{bellemare2013arcade,
  title={The arcade learning environment: An evaluation platform for general agents},
  author={Bellemare, Marc G and Naddaf, Yavar and Veness, Joel and Bowling, Michael},
  journal={Journal of artificial intelligence research},
  volume={47},
  pages={253--279},
  year={2013}
}

@article{brockman2016openai,
  title={Openai gym},
  author={Brockman, Greg and Cheung, Vicki and Pettersson, Ludwig and Schneider, Jonas and Schulman, John and Tang, Jie and Zaremba, Wojciech},
  journal={arXiv preprint arXiv:1606.01540},
  year={2016}
}

\appendix
\section{Appendix}
\label{sec:appendix}

\subsection{Additional Details: Hardware Implementation}

\label{subsec:hardware_extra}
We use four pointers to coordinate the enqueue and dequeue processes in the queue. The pointer \texttt{write\_ptr} selects the next segment address from \texttt{available\_mem\_addrs} to store incoming updates, while \texttt{read\_ptr} retrieves the next segment address from \texttt{out\_mem\_addrs} to fetch outgoing updates.
The pointer \texttt{append\_out\_addr} places the address of each incoming update at the tail of \texttt{out\_mem\_addrs}, and \texttt{append\_available\_addr} stores the next free segment address at the tail of \texttt{available\_mem\_addrs} once the update in that segment is sent out.
These four cyclic registers drive the queue operations.

\noindent\textbf{Handling concurrency.}
In Shesha, we ensure that no more than one update belonging to a given \texttt{Cluster\_ID} is in the queue at any given time.
In the following we provide the corner cases where an incoming update cannot replace or be aggregated with an existing one.
This situation arises when the existing update is locked as it is at the queue head and already scheduled for departure or when parts of it have already exited the queue.
To track such cases and manage cluster-level queuing, \texttt{SheshaQueue} maintains per-cluster metadata using \texttt{cluster\_head} and \texttt{cluster\_tail} circular pointers (cf. Fig.~\ref{fig:overview}).
These pointers identify the positions of the first and second updates in the queue per cluster. The associated \texttt{cluster\_status} register maps each \texttt{cluster\_head} to its queue index, which is then linked to the \texttt{out\_mem\_addr} used to dequeue the corresponding update. Since \texttt{cluster\_head} and \texttt{cluster\_tail} are circular and may wrap around, Shesha uses a three-column \texttt{cluster\_status}. This design ensures that, at any given time, the \texttt{SheshaQueue} can hold and distinguish up to two active updates per cluster, while also detecting when the cluster has no updates in the queue (i.e., when \texttt{cluster\_head} equals \texttt{cluster\_tail}).
\label{subsec:hw_implement_additional}

\noindent\textbf{SheshaQueue Latency Breakdown.}
The OpenNIC shell's interface supports data movement of $512$ bits per clock cycle, and therefore, we restrict each memory block in the \texttt{SheshaQueue} to be $512$ bits ($64$ bytes) in size. To store a standard Ethernet packet of $1500$ bytes, we combine $24$ memory blocks into a memory segment. Given that Shesha operates at a clock frequency of $250$MHz (i.e., $4$ns per cycle), transmitting a $1500$ byte packet requires $96$ns. 

\subsection{Additional Details on the AoM Verification}
\label{subsec:AoM_details}


For simplicity, we assume that the ACK contains the global model update from the PS.
Fig.~\ref{fig:aoi_wave} shows the AoM $\Delta(t)$ at the PS over time $t$, i.e., how old the last model update received at the PS is and hence how old the global model distributed to the asynchronous workers is.
When a model update's arrival time to the engine ($A$) is after the delivery time ($D$) of the previous update (time points $A(1) \dots A(3),D(1)\dots D(3)$ in Fig.~\ref{fig:aoi_wave}) there is no aggregation or replacement.
If the arrival time of the next update is before the delivery time of the previous update, then we see model aggregation or replacement in the queue. 
This can be observed in Fig.~\ref{fig:aoi_wave}, for example, for the update arriving at $A(4)$, which is aggregated or replaced by the update arriving at $A(5)$. 
The figure shows the AoM function in solid red if the aggregation does not occur. In fact, if aggregation does not occur the progress of the AoM function further worsens due to additional queueing, which is not shown in the figure for simplicity. 
Replaced updates are depicted in dashed red, while the aggregation outcome is depicted in dashed black, and the AoM function with the aggregation and replacement is in solid black. 
We can see that the aggregation leads to lower AoM.
Obviously, lower AoM means fresher model updates, such that the updates have collected more experience in the worker-side training and also a less stale global model. 
{\color{\revisioncolor} Notably, this interaction between freshness and aggregation may extend beyond async training, e.g., for aggregating in-network monitoring information.}

\noindent\textbf{Peak AoM.} The peak Age-of-Model $\Delta_p$ is the maximal AoM just before a valid model leaves the accelerator engine, resp. reaches the PS. It is depicted in Fig.~\ref{fig:aoi_wave} as the black dots of the age process. 
The peak AoM $\Delta_p(m)$ after update $m$ (cf. peaks $\Delta_{p}(5)$ and $\Delta_{p}(8)$ in Fig.~\ref{fig:aoi_wave}) is received is 
$\Delta_p(m) = (D(m) - A(m^\prime)) \cdot 1_{\left\{D(m) < A(m+1)\right\}} $
where $m^\prime$ is the index of the latest departed update, $m^\prime = \max \{ i < m: D(i) < A(i+1)\}$, and $1_{\left\{X\right\}}$ is the indicator function. 
The extension of this model for multiple clusters is straightforward. 



\begin{wraptable}{r}{0.49\textwidth}
\vspace{-8pt}
\centering
\caption{AoM model symbols}
\vspace{-10pt}
\label{tab:smt_AoM}
\footnotesize
\setlength{\tabcolsep}{4pt}
\renewcommand{\arraystretch}{1.05}
\begin{tabular}{ll}
\toprule
\textbf{Symbol} & \textbf{Description} \\
\midrule
$\theta$ & output link rate \\
$U$ & total number of clusters \\
$\rho$ & size of each model \\
$A(m)$ & update $m$ arrival time at the engine \\
$D(m)$ & update $m$ departure time from the engine \\
$\Delta_p(m)$ & peak AoM after update $m$ \\
$T_{Q}(m)$ & waiting time of update $m$ in queue \\
$Q(m)$ & queue size when update $m$ arrives \\
$Q_{\max}$ & max queue size \\
$ACK_{Q}(m)$ & queue size when update $m$ departs \\
$\Delta_{w}(m)$ & time since the worker received last ACK \\
\bottomrule
\end{tabular}
\vspace{-8pt}
\end{wraptable}
To formally verify that the chosen worker side transmission control parameters satisfy system-wide constraints we resort to Satisfiability Modulo Theories, a method that enables reasoning about the system behavior using logical constraints over time, queues, and worker side update send rates. The SMT framework can be used to ask queries in form of logical first-order formula with predicates over the AoM processes and deduce corresponding admissible regions for the configuration of the worker-side transmission control functions. 

Table~\ref{tab:smt_AoM} provides all symbols required for the AoM model formulation.
The logical constraints that govern the behavior of the engine are as follows:
\textbf{(i)} Departure time of update $m^u$ (update $m$ of cluster $u$) from the engine, denoted as $D^{u}(m)$, is the sum of the arrival time of the update $A^{u}(m)$ and waiting time of the update in the \texttt{SheshaQueue}, $T_{Q}^{u}(m)$, given the update is not replaced. Hence, $ D^u (m) = (A^u (m) + T_Q^u (m)) 1_{\left\{D^u(m) < A^u(m+1)\right\}}$. 
\textbf{(ii)} The number of updates in the queue when update $m^u$ arrives is the sum of all updates of different clusters that arrived earlier than $m^u$ and will depart after update $m^u$ arrived. Hence, $Q_m^u = \sum_{v^\prime=1}^{U} 1_{\left\{(A^{v^\prime}(m^\prime) < A^v(m)) \land (D^{v^\prime}(m^\prime) > A^v(m))\right\}}$.
\textbf{(iii)} The time required to service one model is $\rho/\theta$, which sets the minimum time difference between any two departures in the queue. 
Hence, $D^u(b) - D^v(a) \geq \rho/\theta$, $\forall a,b, v\neq u, b>a$.
\textbf{(iv)} The queue size when update $m^v$ departs is the sum of all the updates that arrived between the arrival and the departure of $m^v$ and were not replaced or aggregated with a pre-existing update. Hence, the ACK signal for update $m$ which is going back from the engine to the worker carries this queue size and is given by $ACK_{Q,m}^v = \sum_{u=1}^{F} 1_{\left\{ (D^v(m) > A^u(n) > A^v(m)) \land (A^u(n) > D^u(n-1))\right\}}$.
\textbf{(v)} The feedback for update $m^v$ is considered fresh if the time since the last ACK received by cluster $u$ is below the threshold $\Delta_T$. Hence, freshness is defined as $1_{\left\{ \Delta_w^u(m) \leq \Delta_T \right\}}$.
\textbf{(vi)} Based on the freshness, $P_s$ is given in §~\ref{sec:worker_algo}.
\begin{wraptable}{r}{0.5\textwidth}
\vspace{-8pt}
\centering
\caption{\color{\revisioncolor} SMT verification time as the number of clusters sharing a bottleneck increases.}
\vspace{-10pt}
\label{tab:smt_scaling}
\footnotesize
\setlength{\tabcolsep}{4pt}
\renewcommand{\arraystretch}{1.05}
\begin{tabular}{|c|c|c|}
\hline
\textbf{\# Clusters} & \textbf{Verification Time} \\ \hline
2 &  0.125 s \\ \hline
3 &  2.395 s \\ \hline
4 &  39.963 s \\ \hline
5 &  $180$ s \\ \hline
\end{tabular}
\vspace{-10pt}
\vspace{-8pt}
\end{wraptable}

{\color{\revisioncolor} \noindent \textbf{AoM Degradation as an objective and solver run-time evaluation.}
We use the SMT verifier to check if the AoM increase caused by network sharing stays below a given bound. For each cluster $u$, the formal constraints capture the average $\Delta_p$ in the shared setting, denoted as $\overline{\Delta}^{u}_{p,\mathrm{shared}}$, and compare it with the average $\Delta_p$ from a standalone run of the same cluster, denoted as $\overline{\Delta}^{u}_{p,\mathrm{std}}$. We define the normalized AoM degradation of cluster $u$ as
$
G_u =
(
\overline{\Delta}^{u}_{p,\mathrm{shared}} -
\overline{\Delta}^{u}_{p,\mathrm{std}}
)/
\overline{\Delta}^{u}_{p,\mathrm{std}}.
$

The SMT query asks for a bounded trace in which some cluster exceeds a target degradation bound $\Gamma$,i.e.,
$\exists u \quad G_u > \Gamma$ .
Equivalently, the safety condition requires
$
\forall u: 
\overline{\Delta}^{u}_{p,\mathrm{shared}}
\leq
(1+\Gamma)\overline{\Delta}^{u}_{p,\mathrm{std}}$.
We further check the degradation imbalance between clusters using
$|G_u-G_v| \leq \epsilon_{\mathrm{deg}} .
$
This condition is important for heterogeneous DRL workloads, where clusters can have different update-generation rates and therefore different AoM values even when they run alone.

The SMT solver will search for such bounded traces and either prove they do not exist or provide concrete counter examples. We evaluate our SMT-based reasoning framework for competing clusters, base threshold $\bar{\Delta}_T=150$~msec, $\Gamma=0.25$, $\epsilon_{\mathrm{deg}}=0.10$ and $\rho/\theta=2$ (see Tab.~\ref{tab:smt_AoM} for all symbols). We verify correctness under both uniform and non-uniform update generation: (i) when both clusters generate updates every $100$ msec, and (ii) when one cluster sends updates every $100$ msec while the other transmits every $300$ msec.
Table~\ref{tab:smt_scaling} reports the SMT verification time as the number of clusters varies. The solver spends most of this time on pairwise queue-interaction constraints across clusters and updates. We leave further optimization of the SMT verification times for larger, more dynamic data center networks to future work.
}

{\color{\revisioncolor}
\subsection{DRL Workload and Async PPO Workflow Details }


    


\begin{table*}[ht]
\centering
\scriptsize
\setlength{\tabcolsep}{8pt}
\renewcommand{\arraystretch}{0.9}
{\color{\revisioncolor}
\begin{tabular}{|l|c|c|c|}
\hline
\textbf{Parameter} &
\textbf{\texttt{LunarLander-v3}} &
\textbf{\texttt{HalfCheetah-v4}} &
\textbf{\texttt{Atari Pong}} \\
\hline

Rollout per update
& $4$K steps
& $8$K steps
& $16$K frames \\
\hline

Target reward
& $200$
& $3000$
& $18$ \\
\hline

Mini batch size
& $64$
& $128$
& $256$ \\
\hline

PPO epochs
& $10$
& $10$
& $10$ \\
\hline

Learning rate
& $5\times10^{-5}$
& $3\times10^{-4}$
& $2.5\times10^{-4}$ \\
\hline

Architecture
& FC $[1024,1024]$
& FC $[1216,1216]$
& CNN $+$ FC $[640]$ \\
\hline

Update size
& $4.3$ MB
& $6.0$ MB
& $8.4$ MB \\
\hline

\# Packets per update
& $2731$
& $4096$
& $5462$ \\
\hline

Worker compute time
& $1.45$ s
& $1.78$ s
& $2.65$ s \\
\hline

Worker communication time
& $0.30$ s
& $0.22$ s
& $0.45$ s \\
\hline

Horizon Time
& $180$ min
& $180$ min
& $540$ min \\
\hline


\end{tabular}
}
\caption{\color{\revisioncolor}
Workload characterization for the three DRL workloads used in this work. Worker compute time measures rollout collection and local PPO optimization per update. Worker communication time measures serialization, segmentation, and transmission of one update, excluding queueing inside Shesha or the baselines.
}
\vspace{-8mm}
\label{tab:workload_characterization_all}
\end{table*}

We initialize $81$ independent PPO workers for which the hyperparameter values are given in Tab.~\ref{tab:workload_characterization_all}. The workers collect experience and do training independently in their own workload-specific environment. Once a training loop is completed for a particular worker, it sends its updated gradients to the PS. The PS calculates the average of the values it already contains and those it has received, and sends it back to the worker. 
A worker stops training once the reward has converged. 
This continues until the rewards for all workers converge.    }

{\color{\revisioncolor}
\subsection{FPGA Resource Consumption}
\label{subsec:fpga_resources}
We implement aggregation with pipelined FP32 adders. Specifically, the ShehsaQueue processes each $512$-bit AXI beat as $16$ parallel $32$-bit lanes and assigns one FP32 adder to each lane, for a total of 16 FP32 adders per beat. 
\begin{wraptable}{r}{0.5\textwidth}
\vspace{-8pt}
\centering
\caption{\color{\revisioncolor} Resource utilization for Shesha's acceleration engine on Alveo U55C.}
\vspace{-10pt}
\label{tab:util-perc}
\footnotesize
\setlength{\tabcolsep}{4pt}
\renewcommand{\arraystretch}{1.05}
\begin{tabular}{|c|c|c|c|c|}
\hline
\makecell{\textbf{Queue size}\\\textbf{[\# pkts]}} &
\textbf{LUTs} &
\textbf{FFs} &
\textbf{BRAM} &
\textbf{URAM} \\
\hline
770  & 5.04\% & 0.66\% & 0.99\% & 7.52\% \\
\hline
1540 & 5.35\% & 0.84\% & 0.99\% & 15.04\% \\
\hline
2310 & 5.67\% & 1.02\% & 0.99\% & 22.56\% \\
\hline
3080 & 5.99\% & 1.19\% & 0.99\% & 30.08\% \\
\hline
3850 & 6.31\% & 1.37\% & 0.99\% & 37.60\% \\
\hline
\end{tabular}
\vspace{-10pt}
\vspace{-8pt}
\end{wraptable}
We chose this setup because it matches the native beat width of the datapath and allows the design to aggregate an entire beat in parallel instead of serializing the arithmetic across multiple cycles. Tab.~\ref{tab:util-perc} reflects this architecture. The LUT usage remains relatively low and grows only slowly with queue size because the FP32 adder pipelines, aggregation control, and queue-management logic contribute a fixed logic cost that does not depend strongly on the number of packet slots. The FF usage follows the same trend but stays lower because the design uses registers mainly for pipelining and state tracking rather than for bulk storage. The BRAM usage remains nearly constant because the design places only small FIFOs and metadata buffers in BRAM. In contrast, the payload storage dominates the memory footprint, and the design maps that storage to URAM. For this reason, URAM utilization scales almost linearly with queue size, while LUT, FF, and BRAM utilization increase much more gradually. Selective gradient aggregation (\S\ref{sec:live_training}) and querying the number of active clusters (\S\ref{sec:worker_algo}) incur negligible hardware overhead, because the design implements them with simple reward comparisons and lightweight counters.





\begin{figure}[t]
    \centering
    \begin{subfigure}[t]{0.49\textwidth}
        \vspace{0pt}
        \centering
        \caption{}
        \includegraphics[width=\linewidth]{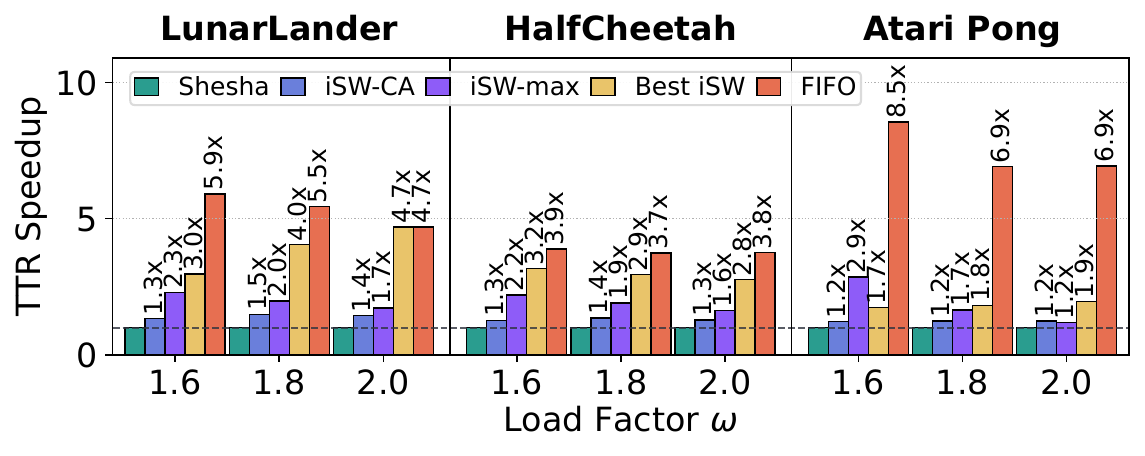}
        \label{fig:normalized_ttr_summary}
    \end{subfigure}
    \hfill
    \begin{subfigure}[t]{0.49\linewidth}
        \centering
        \caption{}
        \includegraphics[width=\linewidth]{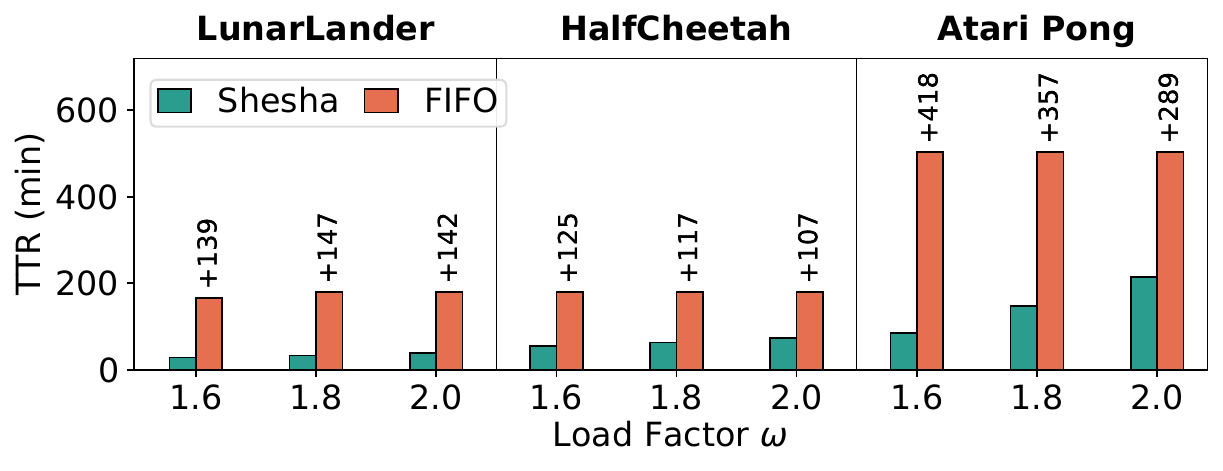}
        \label{fig:fifo_TTR_penalty}
    \end{subfigure}
    \vspace{-9mm}
    \caption{\color{\revisioncolor}(a) TTR Speedup, where Shesha is the reference. Shesha outperforms the baselines in all configurations. (b) Absolute TTR for FIFO. Numbers above bars show the additional TTR in minutes relative to Shesha.}
    \label{fig:worker_reward_and_fifo_penalty}
    \vspace{-2mm}
\end{figure}

{\color{\revisioncolor}
\subsection{Additional Evaluation Results}
\label{sec:isw_congestion_loads}
\textbf{Additional Microbenchmarks (\S\ref{sec:networkbenchmarks}):} Tables \ref{tab:network_isw_isw-ca_out_varied_40}- \ref{tab:network_isw_isw-ca_out_varied_80} show the effect of the batching window size on iSW and iSW-CA for different congestion levels. The setup is the same as in \S\ref{sec:networkbenchmarks}. The input rate is $100$~Gbps, and output rates are $40$ and $80$~Gbps, respectively. Queue depth is half the model size (770 packets). In Tab.~\ref{tab:network_isw_isw-ca_out_varied_40}, we look up the window size ($260 \mu s$) that achieves the same aggregation rate of $56\%$ as Shesha. In comparison, Shesha reduces packet drop to $10.5\%$ and queueing delay of $142.6~\mu s$. Compared to our extension iSW-CA, Shesha reduces the queueing delay by $65.5$\%. In Tab.~\ref{tab:network_isw_isw-ca_out_varied_80}, we look up the window size ($50 \mu s$) that achieves the same aggregation rate of $20\%$ as Shesha. At this operating point, all schemes have the same drop rate. Shesha reduces queueing delay by approximately $44\%$ compared to iSW and iSW-CA.

\begin{table*}[t]
\centering
\scriptsize
\setlength{\tabcolsep}{2pt}
\renewcommand{\arraystretch}{0.92}
\color{\revisioncolor}
\begin{subtable}[t]{0.495\textwidth}
\centering
\caption{\color{\revisioncolor}Input rate $100$~Gbps, output rate $40$~Gbps.}
\label{tab:network_isw_isw-ca_out_varied_40}
\resizebox{\linewidth}{!}{%
\begin{tabular}{|c|cc|cc|cc|cc|}
\hline
\textbf{Window} &
\multicolumn{2}{c|}{\textbf{Drop Rate} (\%)} &
\multicolumn{2}{c|}{\textbf{Queue Delay} ($\mu$s)} &
\multicolumn{2}{c|}{\textbf{Agg. Rate} (\%)} &
\multicolumn{2}{c|}{\textbf{Agg. Size}} \\
\textbf{Size ($\mu$s)} &
\textbf{iSW} &
\textbf{iSW-CA} &
\textbf{iSW} &
\textbf{iSW-CA} &
\textbf{iSW} &
\textbf{iSW-CA} &
\textbf{iSW} &
\textbf{iSW-CA} \\
\hline
10   & 57.7 & 0 & 235.7 & 417.4 & 2.2  & 59.8 & 1.05 & 2.5 \\ \hline
50   & 49.0 & 0 & 270.7 & 420.7 & 10.9 & 60.0 & 1.3  & 2.5 \\ \hline
100  & 38.1 & 0 & 308.4 & 419.3 & 21.8 & 59.4 & 1.54 & 2.4 \\ \hline
260  & 18.6 & 0 & 402.3 & 412.5 & 57.0 & 59.8 & 2.42 & 2.5 \\ \hline
500  & 0    & 0 & 317.7 & 318.1 & 73.2 & 73.7 & 3.7  & 3.8 \\ \hline
1000 & 0    & 0 & 577.6 & 577.8 & 84.5 & 84.3 & 6.5  & 6.7 \\ \hline
\hline
\textbf{Shesha} &
\multicolumn{2}{c|}{10.5} &
\multicolumn{2}{c|}{142.6} &
\multicolumn{2}{c|}{56.0} &
\multicolumn{2}{c|}{2.6} \\
\hline
\end{tabular}%
}
\end{subtable}
\hfill
\begin{subtable}[t]{0.495\textwidth}
\centering
\caption{\color{\revisioncolor}Input rate $100$~Gbps, output rate $80$~Gbps.}
\label{tab:network_isw_isw-ca_out_varied_80}
\resizebox{\linewidth}{!}{%
\begin{tabular}{|c|cc|cc|cc|cc|}
\hline
\textbf{Window} &
\multicolumn{2}{c|}{\textbf{Drop Rate} (\%)} &
\multicolumn{2}{c|}{\textbf{Queue Delay} ($\mu$s)} &
\multicolumn{2}{c|}{\textbf{Agg. Rate} (\%)} &
\multicolumn{2}{c|}{\textbf{Agg. Size}} \\
\textbf{Size ($\mu$s)} &
\textbf{iSW} &
\textbf{iSW-CA} &
\textbf{iSW} &
\textbf{iSW-CA} &
\textbf{iSW} &
\textbf{iSW-CA} &
\textbf{iSW} &
\textbf{iSW-CA} \\
\hline
10   & 15.6 & 0 & 122.5 & 153.9 & 4.3  & 19.9 & 1.05 & 1.3 \\ \hline
50   & 0    & 0 & 66.7  & 68.3  & 21.3 & 21.4 & 1.2  & 1.3 \\ \hline
100  & 0    & 0 & 82.7  & 82.9  & 35.2 & 35.1 & 1.54 & 1.54 \\ \hline
500  & 0    & 0 & 317.7 & 318.1 & 73.6 & 73.7 & 3.7  & 3.8 \\ \hline
1000 & 0    & 0 & 577.6 & 577.8 & 84.5 & 84.3 & 6.5  & 6.7 \\ \hline
\hline
\textbf{Shesha} &
\multicolumn{2}{c|}{0.1} &
\multicolumn{2}{c|}{38.0} &
\multicolumn{2}{c|}{20.0} &
\multicolumn{2}{c|}{1.3} \\
\hline
\end{tabular}%
}
\end{subtable}

\caption{\color{\revisioncolor} Impact of the batching window on iSW and iSW-CA for different output rates.}
\vspace{-15pt}
\label{tab:isw_iswca_window_40_80_side_by_side}
\end{table*}

}
\begin{wrapfigure}{r}{0.49\textwidth}
    \vspace{-3mm}
    \centering
    \includegraphics[width=\linewidth]{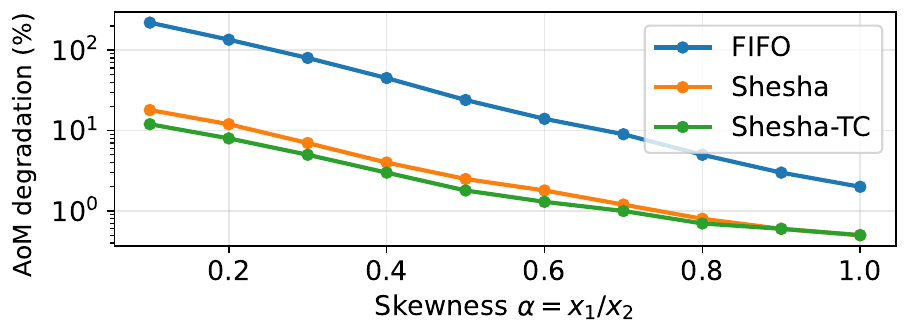}
    \caption{\color{\revisioncolor}AoM degradation for $S1$ versus topology skewness $\alpha=x_1/x_2$, where $x_i$ is the outgoing link capacity of $SW_i$. For $\alpha<1$, congestion is created at $SW_1$, forming a localized bottleneck for cluster group $S1$.}
    \label{fig:aom_deg_s1}
    \vspace{-4mm}
\end{wrapfigure}
\noindent\textbf{Topology skewness (\S\ref{sec:multihop_ns3}).} We start from the setup in \S\ref{sec:multihop_ns3} and study asymmetric link capacities by varying
$\alpha=x_1/x_2$, where $x_1$ and $x_2$ are the outgoing link
capacities of $SW_1$ and $SW_2$. We keep $x_2$ fixed and reduce $x_1$
to create a localized bottleneck for $S1$. As
Fig.~\ref{fig:aom_deg_s1} shows, FIFO sharply increases
the AoM degradation of $S1$ as $\alpha$ decreases. Shesha and
Shesha\_TC keep the degradation low across the entire range of $\alpha$.
The degradation of $S2$ remains much smaller because it is not behind the constrained $SW_1$ side.

}

\end{document}